# Encapsulation theory fundamentals.

Edmund Kirwan[*]
www.EdmundKirwan.com

## ABSTRACT

*This paper proposes a theory of encapsulation, establishing a relationship between encapsulation and information hiding through the concept of potential structural complexity (P.S.C.), the maximum possible number of source code dependencies that can exist between program units in a software system. The P.S.C. of various, simple systems is examined in an attempt to demonstrate how P.S.C. changes as program units are encapsulated among different configurations of subsystems.*

## 1. INTRODUCTION

This paper arguably presents little that is new and rather examines the current practices of encapsulation and information hiding from a shifted perspective. For experienced programmers, this paper perhaps merely explores the adage, "Exploit encapsulation and information hiding optimally." To do this, we must consider complexity.

No consensually-agreed definition of complexity exists in the field of computer science. Lay definitions suggest that complexity is, "A conceptual whole made up of complicated and related parts; complicated in structure; consisting of interconnected parts."

This paper tries to define, not complexity, but a type of complexity and investigates both how computer programs express this type of complexity and how this complexity may be controlled without reducing a program's functionality.

## 2. Foundation

In the seminal work on information hiding [2], Parnas argues that modules should be designed to hide both difficult decisions and decisions that are likely to change. Four aspects of this technique are noted.

Firstly, it is a human that designs modules and it is a human that evaluates whether a decision is difficult or likely to change. Whatever information hiding concerns, therefore, it is practised for the benefit of humans, not machines; in software engineering, this is equivalent to saying that information hiding concerns source code (or models if model-driven development is practised), not compiled/run-time code. A high-level computer language may support information hiding, but once that program is compiled, the resulting executable code need not reflect the information hiding strategies of the original source.

Secondly, the designing-to-hide strategy is essentially one of restriction rather than liberation. The hiding of a decision necessarily entails a reduction of system information from the point of view of some observer.

Thirdly, the target of any proposed restriction is not the decisions being hidden themselves but the relations between modules involving those decisions. Parnas is not advocating the elimination of decisions but the elimination of inter-module relations towards those decisions. As he further clarifies, "The relation we are concerned with is *uses* or *depends upon*."

Fourthly, information hiding concerns the potential as well as the immediate. Once a decision is hidden within a module, it is hidden from both extant observers and observers yet to be created.

Thus we claim: information hiding restricts *potential* source code dependencies, i.e., it restricts source code dependencies that have not yet been implemented.

Consider the system of four program units in figure 1.

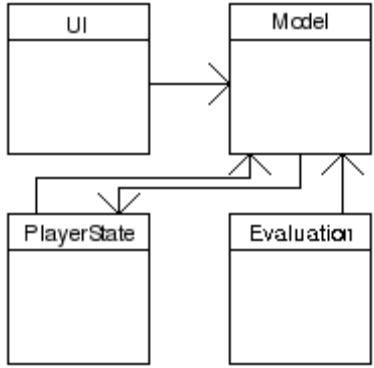

*Figure 1: System of four program units*

Here, both the UI and the Evaluation program units have a dependency towards the Model program unit. The PlayerState and Model program units share two dependencies.

Figure 1 thus shows a system of four dependencies. We do not know the information hiding strategy governing this system but we know that any information hiding strategy will concern potential – and not just actual – source code dependencies.

We may thus ask: how many potential dependencies are possible within this system? What is the maximum possible number of dependencies that these four program units could generate? The answer is shown in figure 2.

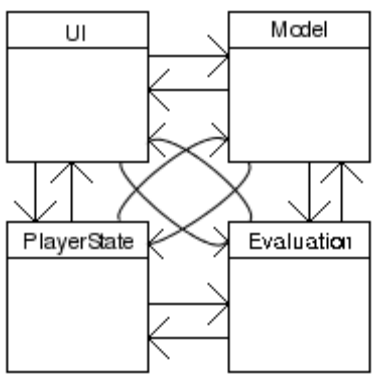

*Figure 2: Maximum interconnectedness.*

A count reveals that the four program units could generate twelve potential dependencies. These twelve dependencies do not, of course, exist in the system, but they could; we say that, independent of the number of actual dependencies the system has, it has twelve potential dependencies.

As mentioned, there is no single definition of complexity in computer science, yet intuitively we might feel that the system (were the dependencies actual) in figure 2 is more complex than that of figure 1, because, all else being equal, figure 2 shows more dependencies.

Let us, therefore, axiomatically define the potential structural complexity (P.S.C.) of a system as the maximum number of dependencies possible between program units of that system.

We thus say that the system shown in figure 1 has a P.S.C. of 12.

### 3. Encapsulation and information hiding

Before examining P.S.C. further, we must define the two key concepts on which it is founded: encapsulation and information hiding. The definitions proposed here may deviate from those more generally available but a definition is only as good as the utility to which it gives rise; the judgement of such utility is left to the student.

These definitions will be introduced, furthermore, in a stepwise manner: beginning quite informally so as to guide the student through familiar terminology, and concluding with a formal definition used for rigorous proofs.

Let us provisionally define, "Encapsulation," as the the placing of program units within a subsystem.

Let us provisionally define, "Information hiding," as the restricting of forming dependencies on any particular program unit within a subsystem from outside that subsystem.

It may be argued that encapsulation and information hiding are orthogonal issues, and that information hiding demands only that *some* separator exists which differentiates those clients that may and may not observe some particular information; there exists no requirement that the separator and the subsystem boundary be one and the same. Usually, however, the subsystem boundary is the separator that demarcates information hiding, so we shall consider only this case.

According to the definitions above, two program units within the same subsystem cannot be information hidden from one another.

## 4. A preliminary appraisal

Returning to figure 2, we have seen that the four, unencapsulated program units generate twelve dependencies. This is an example of an unencapsulated system, in that the four program units are not separated into subsystems. In fact, given any unencapsulated system, $G$, of $n$ program units, the potential structural complexity $s(G)$ is given by:

$$s(G) = n^2 - n$$

(See theorem 1, though this equation has long been known in graph theory.) In our example, $n = 4$, and we see that:

$$4^2 - 4 = 16 - 4 = 12$$

It is clear, then, that the P.S.C. of an unencapsulated system is proportional to the square of the number of its program units, or that the P.S.C. rises quadratically with the number of program units involved. This is also the upper ceiling of P.S.C. for any given $n$ program units: it is impossible for any system to generate a higher P.S.C. than this.

Here we gain our first insight into the nature of potential structural complexity: for unencapsulated systems, P.S.C. scales quadratically as program units are added. Figure 3 shows the P.S.C. for unencapsulated systems of one up to one hundred program units, it is thus said to show the P.S.C. response of these systems.

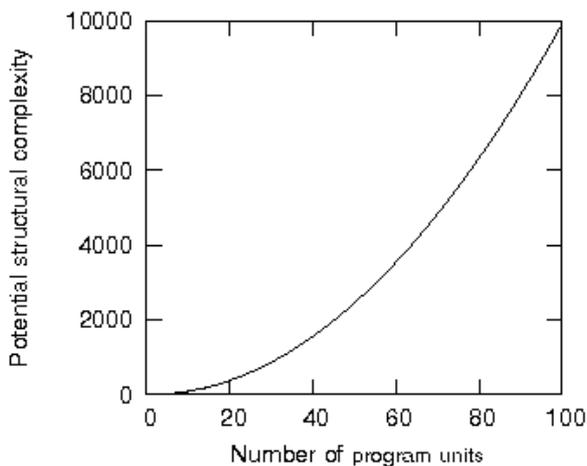

*Figure 3: Potential structural complexity as a function of the number of program units in an unencapsulated system*

Potential structural complexity for unencapsulated systems, as we can see from figure 3, scales more than linearly. In general, however, any complexity response rising faster than linearly threatens to make large systems disproportionately more difficult to create than small systems. We may then wonder what an ideal P.S.C. response for a system might be.

Such an ideal response would show a reduction of P.S.C. with the addition of program units, yet this is clearly impossible, as P.S.C. measures the maximal possible connectedness of the program units of a system, and each new program unit must by definition involve the addition of at least one dependency within the system or else the newly added program unit cannot meaningfully be said to be part of that system.

A more realistic, desired response, therefore, would entail adding as little P.S.C. as possible with the addition of program units. The rest of this paper is concerned with the search for such a desired response. We shall do so by performing some simple experiments within three encapsulation contexts.

## 5. Encapsulation contexts and experiment

An encapsulation context describes the environment in which encapsulation is employed, classifying this encapsulation based on the nature of its dependency restriction. The first encapsulation context examined is non-hierarchical encapsulation; this context ignores, from the point of view of dependency formation, the recursive declaration of subsystems within subsystems, treating all subsystems as though they were declared in isolation from one another.

This paper focuses on this context; the other two encapsulation contexts we shall review – in far less detail – are the one-dimensional hierarchical encapsulation context and the two-dimensional hierarchical encapsulation context, both defined later in the paper.

To explore the potential structural complexity of systems residing with the non-hierarchical encapsulation context, we shall perform the three experiments.

The first experiment is called, "The fixed-system experiment." This experiment takes a fixed number of program units and a fixed number of subsystems, and examines how the P.S.C. of the system changes as program units are distributed over the subsystems.

The second experiment is called, "The varied-region experiment," which takes a fixed number of program units and examines how the P.S.C. of the system changes as program units are uniformly distributed over an increasing number of subsystems, that is, as the system moves from one uniform distribution to another.

The third experiment is called, "The system-growth experiment," which takes a number of systems composed of increasing numbers of program units and examines how the minimum P.S.C. of the configuration increases as more program units are added.

(We have, indeed, already performed one such system-growth experiment: figure 3 showed the results of the this experiment for an unencapsulated system, the graph showing the quadratic nature of the P.S.C. response of the first 100 unencapsulated systems.)

## 6. Non-hierarchical encapsulation

Non-hierarchical encapsulation means that the information hiding of a subsystem is independent of its relationship with all other subsystems (a point to which we shall return in section 6.7). A system may be hierarchically structured yet not be hierarchically encapsulated (in fact Java systems, in advance of the introduction of super-packages, are usually hierarchically structured but not hierarchically encapsulated).

Take the system described in figure 4, which shows four subsystems: *a, b, c* and *d*, with *c* declared within *b*.

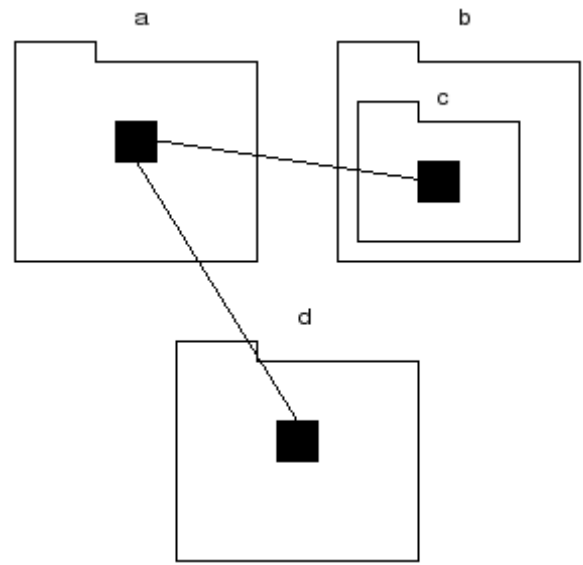

*Figure 4: Recursive/nested structure*

Figure 4 shows a hierarchically structured system, as *c* is contained within *b*, but in a non-hierarchically encapsulated system, such structure has no effect the ability of program units to form dependencies. In the diagram, the program unit in *a* can form dependencies on the program unit in *c* just as it can on the program unit in *d*.

### 6.1. The fixed-system experiment

This experiment will attempt to answer the question: is the potential structural complexity of a system minimised when it is uniformly distributed, that is, when its program units are uniformly distributed over its over subsystems?

Consider four program units and two subsystems. We shall examine all possible configurations in which these four program units can be distributed over the two subsystems.

We must first, however, chose an information hiding strategy that will remain constant throughout the experiment. Our information hiding strategy will simply state the maximum number of program units that are visible outside a subsystem (as we shall see later, this is achieved in Java by declaring a class `public`). We shall chose to have only one program unit per subsystem visible outside that subsystem; that is, all but one program unit will be information-hidden within each subsystem. The program units that are information-hidden will be coloured white; the other program units (the, "Public," program units) will be coloured black.

Let us begin by putting all four program units in one subsystem while leaving the other subsystem empty. See figure 5 for the resulting P.S.C. and note that, for pictorial convenience, the arrows indicating dependency directionality have been omitted.

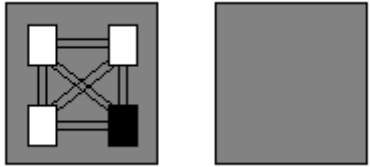

*Figure 5: Four program units in 1 subsystem.*

Here we see twelve dependencies between the four program units; the P.S.C. is twelve, just as it was when the program units were unencapsulated.

Next, we shall examine the configuration whereby three program units reside in one subsystem and one in the other, see figure 6.

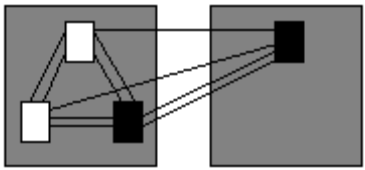

*Figure 6: Three program units in one subsystem.*

Here, we see ten dependencies; the P.S.C. is ten, two fewer than in figure 5. Why are there fewer dependencies than in figure 5?

The reason there are fewer dependencies concerns the two white program units on the left. As before, they are information hidden within their subsystem, but in figure 5, this information hiding is irrelevant, as all program units within a subsystem are visible to one another and thus can form dependencies towards one another; only program units outside the subsystem are restricted from forming dependencies towards information hidden program units.

In figure 6, however, the program unit on the right now cannot form dependencies towards those information-hidden program units in the subsystem on the left.

It is because information hiding has come into play in figure 6, restricting one program unit's ability to form dependencies towards two others, that the P.S.C. has fallen by two.

Let us proceed and examine the system with the four program units uniformly distributed over two subsystems, see figure 7.

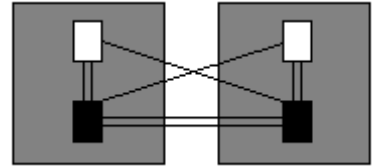

*Figure 7: Two program units in each subsystem.*

Now, there are only eight dependencies; the configuration in figure 7 has a P.S.C. of eight, two fewer than in figure 6, and a full four dependencies fewer than the original figure 5.

Again, the reason for this decline in P.S.C. concerns information hiding. Some scrutiny of the evolving configurations reveals that the collapse of a program unit's ability to form bi-directional dependencies with others within its own subsystem has been unmatched by the rise in its ability to form unidirectional dependencies towards the visible program units of the other subsystem.

In figure 5, each program unit can form bidirectional dependencies towards three others; in figure 6, each program unit can form bidirectional dependencies towards only two others; in figure 7, each program unit could can form bidirectional dependencies towards only one other program unit.

This asymmetry between forming bidirectional dependencies within a subsystem and unidirectional dependencies without is at the heart of encapsulation theory and we shall now further refine P.S.C. to reflect it.

We shall define the *internal* potential structural complexity as the number of potential dependencies formed by program units on one another within a subsystem, and we shall define the *external* potential structural complexity as the number of potential dependencies formed by program units towards program units outside their subsystem. The P.S.C. of a subsystem is then the sum of these two terms.

Our earlier scrutiny can now be re-phrased as follows: the fall in the system's overall P.S.C. stems from the collapse of the internal P.S.C.'s being unmatched by the rise in external P.S.C..

Before looking deeper into this

relationship, we should conclude the experiment, though it turns out that there is little more to learn as continuing to move program units into the subsystem on the right immediately confronts us with mirror-images of earlier configurations; figure 8, for example, shows one program unit in the subsystem on the left and three in the subsystem on the right.

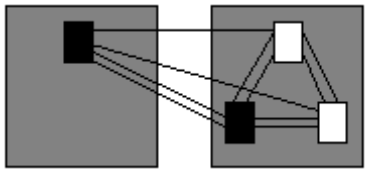

*Figure 8: Three program units in one subsystem.*

Figure 8 has the same P.S.C. as that of figure 6 (reflecting the images has no effect on the number of dependencies formable). Figure 7, therefore, exhibits the lowest P.S.C. of all the configurations. Recall that the question to be answered by this experiment was: is the P.S.C. of a system minimised when it is uniformly distributed? The answer appears to be yes.

The P.S.C. equations governing the systém reflect this answer.

The equation for the internal P.S.C. of a subsystem takes the same form as that already encountered for an unencapsulated system, namely, where $K$ is subsystem of $n$ program units, then the internal P.S.C., $s_{in}(K)$, is (loosely):

$$s_{in}(K) = n(n-1) = n^2 - n$$

(See theorem 1.2.) The external P.S.C. of a subsystem is less elegant, but given here for completeness:

$$s_{ex}(K_i) = \left|K_i\right|\left(\left|h(G)\right| - \left|h(K_i)\right|\right)$$

(See theorem 1.4.) Although not immediately apparent, there is a difference between the two equations in terms of order. The equation for external P.S.C. contains no powers, that is, it is linear. This can be seen more clearly by noting that the three terms on the right-hand-side of the equation are all constants, thus by arbitrarily assigning constant values to these terms, we can re-write the equation as:

$$s_{ex}(K_i) = a(b-c)$$

All the powers on the right-hand-side can thus be seen to be 1; there are no $a^2$ or $c^3$ terms, for example.

This is why uniform distribution minimises P.S.C.: because the quadratic nature of internal P.S.C. will tend to dominate the linear nature of external P.S.C. whenever program units cluster unevenly into subsystems.

So, we have tentatively concluded that a system's P.S.C. is minimised by its being uniformly distributed. When examining four program units, it is clear that having two program units in two subsystems is the only uniform distribution available; but what if there are many possible uniform distributions?

What if we want to encapsulate, for example, twelve program units to minimise P.S.C.: which uniform distribution is best? Six program units in two subsystems? Four program units in three subsystems?

Indeed, is there any configuration that minimises P.S.C.?

## 6.2. The varied-region experiment

Let us perform our second experiment, the aim of which is to investigate whether there exists a configuration that minimises P.S.C., that is, whether the P.S.C. of a specific system is minimised by that system's being in a specific uniform distribution.

Consider a system of twelve program units. We shall divide this system up into an increasing number of subsystems such that an equal number of program units will be in each subsystem at any given time. We shall, again, enforce the information hiding strategy of having only one program unit per subsystem visible outside that subsystem.

First, we shall examine a system that is unencapsulated, see figure 9. Again for pictorial convienence, the remaining figures of this paper will represent both a unidirectional and a bidirectional dependency with just one line.

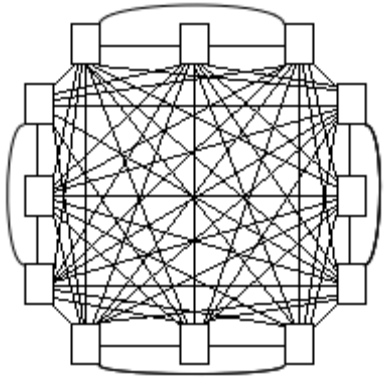

*Figure 9: Twelve program units unencapsulated.*

The unencapsulated system in figure 9 has a P.S.C. of 132. This is as predicted by our previously established equation for the P.S.C. of an unencapsulated system:

$$s(G) = n^2 - n = 12^2 - 12 = 144 - 12 = 132$$

We shall now view the system encapsulated into two subsystems of six program units each, thus there will be just two program units (coloured black) in the system that are visible to all else, whereas five program units in each subsystem will be visible only to one another. See figure 10.

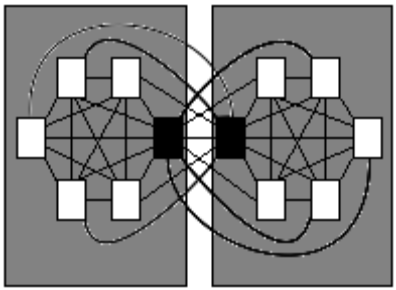

*Figure 10: Two subsystems, six program units in each.*

With the twelve program units encapsulated in two subsystems, the P.S.C. falls to just 72.

Let us make an incautious hypothesis at this point: we shall predict that the P.S.C. falls as the number of subsystems increases. Figure 11 shows the twelve program units encapsulated in three subsystems of four program units each.

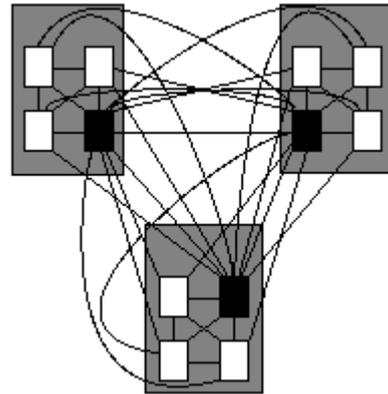

*Figure 11: Three subsystems, four program units in each.*

The P.S.C. of figure 11 is 60, in line with our hypothesis so far: the P.S.C. has fallen as we increase the number of subsystems. Next we shall look at the twelve program units uniformly encapsulated in four subsystems, see figure 12.

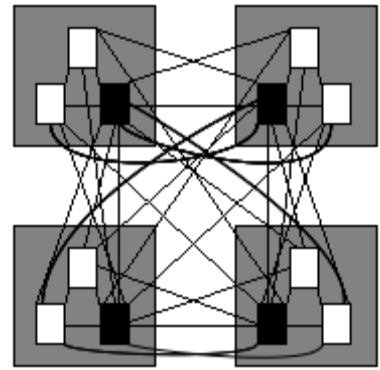

*Figure 12: Three subsystems, four program units in each.*

Here, our hypothesis falters slightly, as the P.S.C. of figure 6 is 60, the same as the P.S.C. of the twelve program units encapsulated into three subsystems. We shall continue the experiment regardless. Figure 13 shows the twelve program units encapsulated in six subsystems.

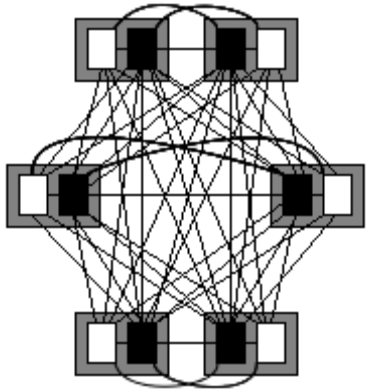

*Figure 13: Six subsystems, two program units in each.*

With six subsystems, the potential structural complexity rises to 72 and our naive hypothesis disintegrates. It seems that increasing the number of subsystems does not automatically decrease the minimum P.S.C. of a system. Figure 14, finally, shows each program unit alone in its own subsystem.

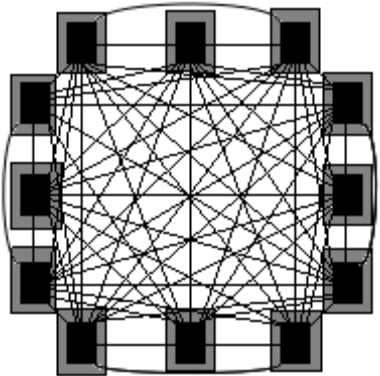

*Figure 14: Twelve subsystems, one program unit in each.*

In figure 14, the P.S.C. has risen to the original value of 132. The results of this experiment are shown most clearly by drawing a graph of the P.S.C. against the number of subsystems into which the twelve program units were uniformly encapsulated, see figure 15.

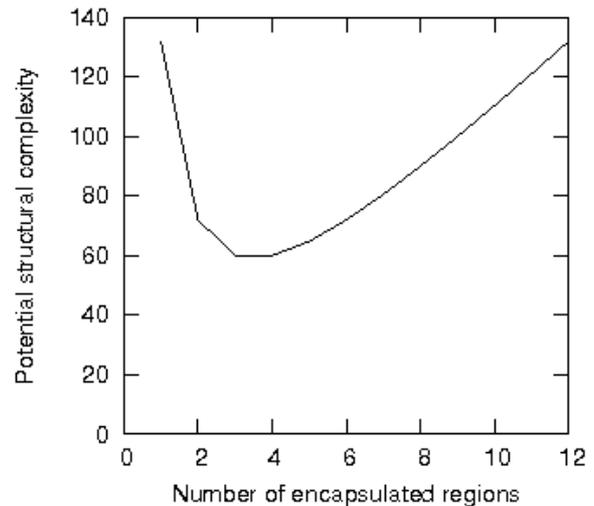

*Figure 15: Potential structural complexity versus number of subsystems.*

With figure 15, then, we seemed to have answered our question: does there exist a system configuration that minimises P.S.C.? From the graph, it appears that both three and four subsystems minimise the P.S.C. of this system.

In fact, for any system in a non-hierarchical encapsulated context with $n$ program units equally distributed over all subsystems and with the same number of externally visible program units, $p$, in each subsystem, it turns out that the number of subsystems, $r_{min}$, at which the P.S.C. is minimised given by the second law of encapsulation:

$$r_{min} = \sqrt{\frac{n}{p}}$$

(See theorem 1.12.) Thus, for 12 program units ($n=12$) each with one program unit externally visible ($p=1$), $r_{min} = \sqrt{12} = 3.46$, though, obviously, we can only have whole numbers of subsystems in any real system.

### 6.3. The system-growth experiment

Our third and final experiment for the non-hierarchical encapsulation context is to examine how the P.S.C. of the P.S.C.-minimised configuration increases as more program units are added.

We shall examine all systems composed of one to one hundred program units. For each system, we shall find the minimum P.S.C. (see theorem 1.14) and plot this minimum P.S.C. as a function of the number of program units.

Figure 16 shows the resulting graph; on

this graph we shall also plot the same one hundred systems whose program units are unencapsulated (that is, the quadratic response already shown in figure 3). In figure 16, the P.S.C. of the unencapsulated systems is the higher line, whereas the P.S.C. of the non-hierarchical encapsulation context is the lower line.

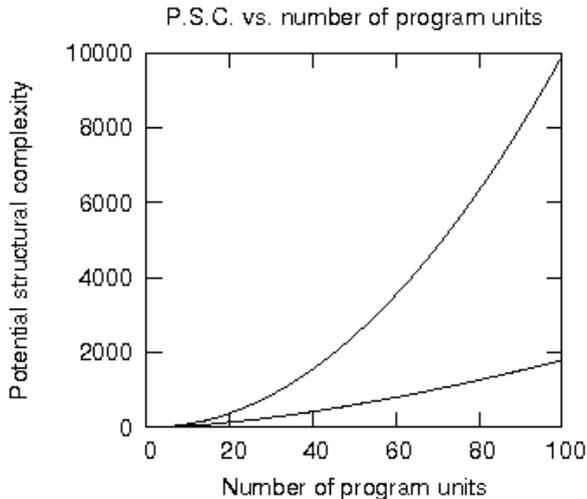

*Figure 16. Unencapsulated and non-hierarchical encapsulation context P.S.C. as a function of the number of program units*

Figure 16 shows the degree to which encapsulation can reduce the P.S.C. of a system: for a system of just one hundred program units, an unencapsulated system expresses over five times more P.S.C. than an encapsulated system.

Examining of the equation of theorem 1.14 reveals that minimised P.S.C. is proportional to $n^{\frac{3}{2}}$ or that the order of the equation is $O(1.5)$. Recall that the P.S.C. equation for an unencapsulated system was quadratic, or of order $O(2)$. This reduction of order $0.5$ is responsible for the differences between the two graphs in figure 16.

## 6.4. The regional information hiding violation

It's perhaps worth examining the role of $p$ in the above equations in more detail. This variable represents the the number of program units within a subsystem that are visible outside that subsystem. In Java, for example, a class declared with the access modifier `public` is visible outside the package in which it resides:

```
package com.rail;

    public class Train {
```

```
    ...
}
```

Here the `Train` class is visible outside the `com.rail` package. We shall define such visibility to be a *regional information hiding violation*, and the variable $p$ above is merely the number of information-hiding violations of a subsystem (indeed the variable, $p$, was chosen as it is the first letter of the word, "Public"). If `Train` is the only class defined as `public` within the `com.rail` package, then the number of information-hiding violations of the `com.rail` package is 1 ($p=1$). If there are two classes defined as `public` in `com.rail`, then the number of information-hiding violations of the `com.rail` package is 2 ($p=2$), etc..

Figure 15 presented the results of the varied-region experiment, showing the P.S.C. of 12 program units as the system moved from one uniform distribution to another. A general equation for this graph can be derived (see theorem 1.8) which calculates the P.S.C. as a function of the number of program units, subsystems and number of regional information-hiding violations. Figure 17 plots this equation by showing, for example, the P.S.C. as 100 program units are distributed over a number of subsystems (note the similarity with figure 15), again with one information-hiding violation per subsystem ($p=1$).

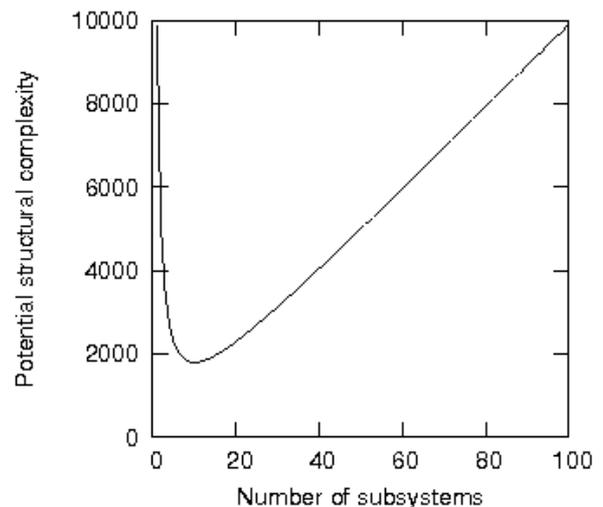

*Figure 17: Potential structural complexity of 100 program units versus number of subsystems*

It may also be seen from the graph that the number of subsystems at which the P.S.C. is minimised is, again, given by the second law:

$$r_{min} = \sqrt{\frac{n}{p}} = \sqrt{\frac{100}{1}} = 10$$

We shall now examine what happens when there are two information-hiding violations per subsystem *(p=2)*, and then three *(p=3)* and then four *(p=4)*; see figure 18.

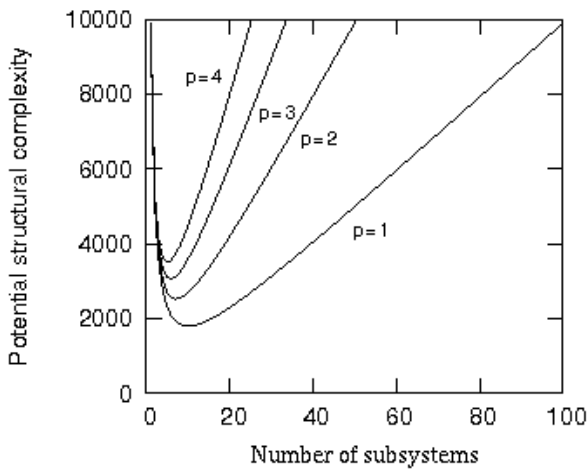

*Figure 18: Potential structural complexity of 100 program units versus number of subsystems with increasing number of public program units per subsystem*

Two details are noteworthy.

Firstly, as the number of information-hiding violations per subsystem rises, the minimum possible P.S.C. also rises. This means that the more `public` program units reside in the system, the higher the minimum P.S.C. will be.

Secondly, as the number of information-hiding violations per subsystem rises, the number of possible subsystems over which the program units can be distributed falls.

This becomes obvious when we consider that, with a requirement of one information-hiding violation per subsystem, then the minimum number of program units per subsystem is one, and thus the system can be distributed over a maximum of 100 subsystems. With a requirement of two information-hiding violations per subsystem *(p=2)*, then the minimum number of program units per subsystem is two, and thus the system can be distributed over a maximum of 50 subsystems.

## 6.5. Real software systems

### 6.5.1. Minimum P.S.C. for p=1

Let us define equivalence by saying that two systems are equivalent if they have the same number of program units and subsystems and they have the same regional information hiding violation.

The systems studied to this point have generally been uniformly distributed because a uniformly distributed system with *p=1* has a lower P.S.C. than any equivalent, non-uniformly distributed systém (see theorem 1.11). Real systems, however, tend not to have only one program unit public per subsystem, and their subsystems usually do not all contain the same number of program units.

Yet even real systems must obey the laws of encapsulation in that they are not free to express any P.S.C. without check; real systems may not express a P.S.C. outside the range established by the first law of encapsulation and theorem 1.14 (with *p=1*).

To visualise this, let us take 1000 randomly generated systems. The creation of these systems will be subject to three constraints: (i) each system must have exactly 100 program units; (ii) each system must contain a random number of subsystems, between 1 and 100; and (iii) each subsystem must contain a random number of public and private program units (though each must have at least one public program unit).

So, for example, a system of three subsystems – each having 12, 7 and 53 private program units respectively and each having 5, 2 and 21 public program units respectively – would be a valid system; a system of 57 subsystems would be a valid system, and so on.

Let us also take the plot of the minimum-possible P.S.C. of a system of 100 program units uniformly distributed, as shown in figure 17, and superimpose on it our 1000 random system configurations, see figure 19, where each cross represents the P.S.C. response of one random system.

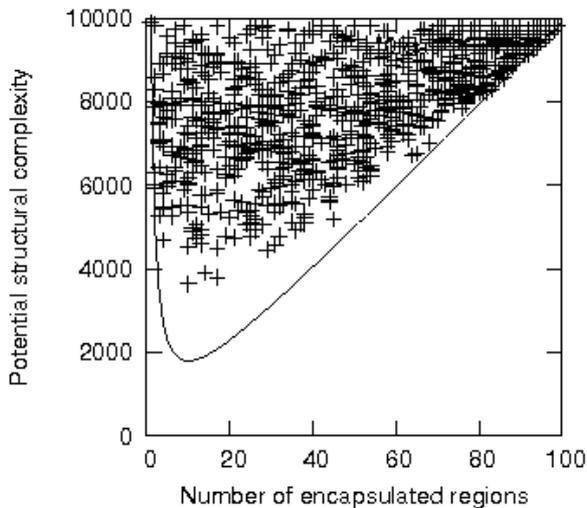

*Figure 19: the P.S.C. of 1000 random systems*

|  | # private program units | # public program units |
|---|---|---|
| Subsystem 1 | 33 | 12 |
| Subsystem 2 | 5 | 50 |

*Table 1: The first A.M.C.*

Table 1 shows a system of 100 program units (*n=100*), two subsystems (*r=2*) and a regional information hiding violation of 31 (*p=31*). The P.S.C. of this system (see theorems 1.2 and 1.4) is 7860. The minimum P.S.C. of the equivalent, uniformly distributed system (see theorem 1.14) is 7935.

We must resort to the conjecture that, for non-trivial systems, encapsulation theory predicts a minimum P.S.C. that lies within 1% of the true minimum. Although non-rigorous, this conjecture is not without some evidence.

Firstly, A.M.C.s appear to become rare very quickly with increasing numbers of program units and subsystems. For a system of two subsystems and 100 program units, brute-force simulations show that A.M.C.s account for 13% of all possible configurations. For a system of five subsystems with the same number of program units, A.M.C.s account for just 0.004% of all possible configurations. etc. See figure 20.

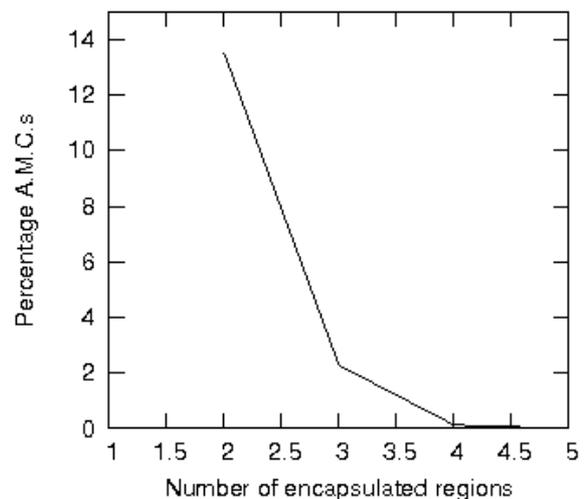

*Figure 20. Percentage A.M.C.s with increasing subsystem*

As can be seen from figure 19, none of the 1000 systems breaches the gradient mandated by the potential structural equation (see theorem 1.8). Here, we may interpret the P.S.C. gradient not as a line but as an edge sweeping out an area – the potential structural complexity surface – onto which every system must map to a single point (and over which each system will trace a trajectory as it develops).

Thus is illuminated for the programmer some real constraints to which programs are (and, indeed, always have been) subjected.

As we shall see, it is theorem 1.11 that provides the connection between real and theoretical systems, allowing us to use the mathematical tools of uniformly distributed systems to gain insight into the non-uniformly distributed systems of real software.

### 6.5.2. Minimum P.S.C. for p>1

Most of the randomly generated systems in figure 19, however, have *p>1*. It would be satisfying if it were proven that the minimum P.S.C. of a uniformly distributed system of *p=2*, for example, was the also lowest possible P.S.C. of the equivalent, non-uniformly distributed systems. This, however, is not the case.

For *p>1*, there exist non-uniformly distributed systems – called anomalous minimised configurations (A.M.C.s) – whose P.S.C. is less than that of their equivalent uniformly distributed systems. Take, for example, the A.M.C. in table 1.

Secondly, if we measure the difference between the theoretical minimum P.S.C. and the P.S.C. of the lowest A.M.C. we arrive at a measure for how accurate the theory is. For a system of two subsystems with 100 program units, encapsulation theory's minimum P.S.C. predicted is within 4% of that of the lowest A.M.C.. For a system of five subsystems, the prediction is within 0.4%. etc. See figure 21.

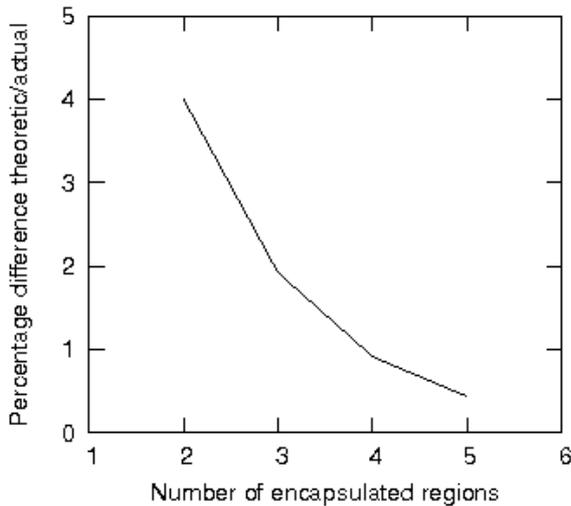

*Figure 21. Percentage distance between theoretical and actual minimum P.S.C.*

Our conjecture is thus based on the extrapolations of both graphs for increasing numbers of program units and subsystems: it appears that, as systems grow, they contain proportionately fewer A.M.C.s, and the minimum P.S.C. of those A.M.C.s is increasingly close to the theoretical prediction. We say that, with *p>1*, for non-trivial systems (*r>5*), encapsulation theory predicts an approximation of the minimum P.S.C..

### 6.5.3. Instability

Figure 18 also reveals a noticeable gap – "Boltzmann's hammock" –  between the low P.S.C. gradient and the nearest clusters of P.S.C. responses. There are very few ways in which a system can be configured such that it is uniformly distributed with a low regional information hiding violation; there are vastly more ways in which it can be configured such that it is non-uniformly distributed and/or does not have a low regional information hiding violation.

This implies that such uniformly distributed, low regional information hiding violation systems are highly unlikely to be achieved through random selection, and thus concentrations of random systém P.S.C. responses peter out as they approach the P.S.C. gradient. (Such statistically unlikely configurations are said to exhibit low entropy.)

Furthermore, given a P.S.C.-minimised

system, any ad hoc updates to that system will tend to increase the P.S.C. of that system, again because of the statistical likeliness that those updates will make the systém less uniformly distributed or give it more regional information hiding violations, thus taking the system further from the P.S.C. minimum. Programmers, of course, do not make ad hoc updates per se, but in this instance, ad hoc simply means, "Without consideration of distribution uniformity and information hiding."

In other words, contrary to the intuitive notion of the desirability of stability, low-P.S.C. systems are in highly unstable configuration equilibria: they reside far from stable configuration equilibria. We shall shortly explore where such a stable equilibrium might lie.

### 6.5.4. Configuration efficiency

Another aspect of encapsulation theory's view of real-world systems concerns program functionality. Just as an all-purpose definition of complexity eludes us, unfortunately so too does an all-purpose definition of functionality: functionality is loosely considered a measure of how much a program does; the more a program does, the more functionality it has. It is also – again loosely – the case that the more program units a system has the more functionality it delivers. We can certainly say that the addition of program units *may* increase a system's functionality.

The P.S.C. equation basically expresses P.S.C. as a function of (among other things) the number of program units in a system, thus it can be viewed as an equation of conversion of P.S.C. to functionality and vice versa. The history of software engineering has largely been that of creating increasingly large programs and then managing the ensuing, increasingly large complexity. The P.S.C. equation, however, offers us a means of turning this history on its head and asking: given a certain amount of potential structural complexity, how much functionality can we efficiently extract?[1]

It is this efficiency that is of interest. Recall that the typical minimum-P.S.C. graph shown in figure 17 shows a system with 100

---

1   Rearranging the terms of the equation in theorem 1.14 gives precise values.

program units differing only in how those program units are encapsulated into subsystems: if functionality is a function of the number of program units, then there will be no change in the functionality of a system as the same 100 program units are re-distributed over subsystems; and yet the second law of encapsulation shows that there is a minimum P.S.C. for any systém (A.M.C.s notwithstanding), regardless of whether the system's expressed P.S.C. approaches this minimum.

Combining these two ideas leads to the proposal that any P.S.C. above the minimum is a rise[2] in a form of system complexity without a necessary rise in concomitant functionality.

This leads to the concept of configuration inefficiency, which shall define as that proportion of P.S.C. a system expresses over and above its minimum possible P.S.C.; the point being that all inefficient P.S.C. is reducible, in theory, to zero. There are two ways of doing this: either by reducing the system's overall P.S.C. through re-encapsulation; or by maintaining the system's overall P.S.C. but adding program units (and thus lowering the inefficient P.S.C. by increasing functionality, a process that may continue until the inefficient P.S.C. is zero, after which the addition of program units must cause the overall P.S.C. to rise again).

The equation for configuration inefficiency $c_i$ is thus:

$$c_i = \frac{s(G) - s_{min}(G)}{s_{max}(G) - s_{min}(G)}$$

Where:

$s(G)$ = actual system P.S.C.

$s_{min}(G)$ = minimum system P.S.C.

$s_{max}(G)$ = maximum system P.S.C.

Though it is more usual to express this as an efficiency $c_e$ which is simply:

$$c_e = 1 - \frac{s(G) - s_{min}(G)}{s_{max}(G) - s_{min}(G)}$$

2 By no means does this paper suggest that all P.S.C. rises above the minumum are to be avoided: there are many other reasons for P.S.C. rises besides even the loose definition of functionality given here.

Thus the configuration efficiency ranges from 0 to 1. A configuration efficiency of 1 means that the system is expressing its minimum possible P.S.C. (all of its P.S.C. is converted to functionality); a configuration efficiency of 0 means that the system is expressing the maximum possible P.S.C. (its P.S.C. is converted to functionality with minimum efficiency).

In figure 18, 1000 systems (each with 100 program units) were randomly generated; we can calculate the configuration efficiency of each of these 1000 systems and then take the average, giving us the average configuration efficiency of a random system of 100 program units. We can then expand this technique and find the average configuration efficiency of all random systems of between 1 and 10,000 program units; the results of this simulation are shown in figure 22.

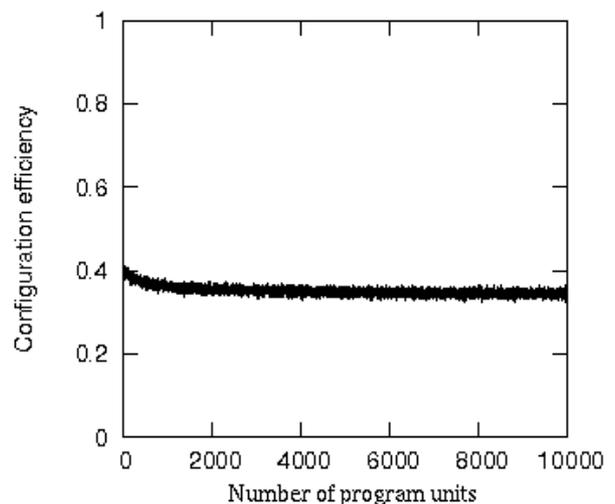

*Figure 22: Average configuration efficiency of random systems*

Figure 22 shows that the average configuration efficiency for a randomly created system is approximately scale-invariant at 0.35. This process is sensitive to the conditions of the system creations, nevertheless this does suggest a tendency for randomness to be associated with a relatively low configuration efficiency.

To investigate further, ten randomly generated systems were subjected to a long series of ad hoc updates; these updates preserved subsystem number but either added or removed program units, or moved program units between subsystems. The results of this investigation are shown in figure 23.

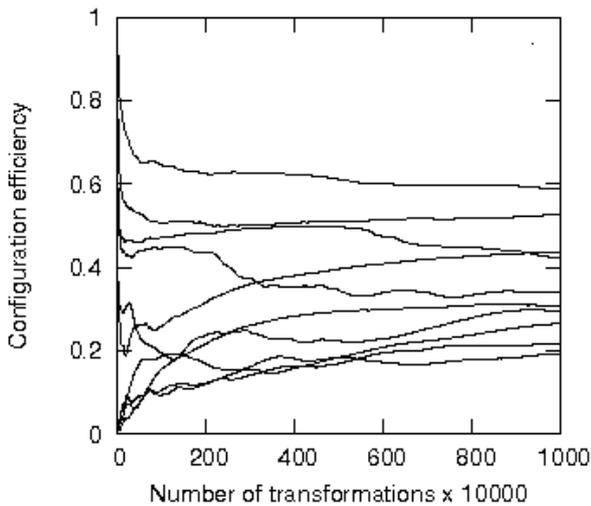

*Figure 23: The configuration efficiency of ten random systems, updated ad hoc.*

Figure 23 shows that a series of ad hoc updates to any system will tend to bring its configuration efficiency to between 0.2 and 0.6, and once a system achieves this belt, subsequent ad hoc updates will tend to cause it to remain within this belt: this thus represents an extremely stable configuration of a system, the system's stable equilibrium. (Such statistically likely configurations are said to exhibit maximum entropy.) We shall define the 0.2-0.6 configuration efficiency range as the stable equilibrium belt.

Real software systems residing below the stable equilibrium belt will tend to increase their configuration efficiency by randomly re-organising all their program units into a random number of subsystems; in other words, as we would expect most encapsulation and information hiding strategies to be more effective at reducing P.S.C. than random selection, we would expect most software systems to reside above the stable equilibrium belt, that is, to express a configuration efficiency of greater than 0.6. We shall investigate this expectation shortly by examining some real software systems.

Encapsulation theory suggests a raft of other software metrics – not least the P.S.C. itself – yet just one is mentioned here: the percentage information hiding violation (I.H.V.), the number of public program units divided by the total number of program units, multiplied by 100. Given that this is, in some sense, a crude measurement of a system's information hiding, this metric offers an excellent, "First appraisal," of a system; a high percentage I.H.V. usually obviates the need to check the configuration efficiency.

Let us examine some open-source, Java software systems from sourceforge [6]. The projects will be taken more or less at random though Jboss and Eclipse are included specifically for their size.

We shall parse them at the third-graph (i.e., class/package level, see section 6.6) with a free, downloadable tool, *Fractality: Free Edition [5]*, which computes a range of encapsulation theory metrics; see table 2 (where, "C.E.," stands for, "Configuration efficiency.") Note that (i) the systems were parsed in their entirety, no attempt was made to separate dedicated source code from included, third-party libraries; (ii) the systems are presumed to reside within non-hierarchical encapsulation contexts; (iii) the source code is presumed to be human-manipulated (not, for example, generated from human-manipulated models, see introduction).

|  | Num. nodes | P.S.C. | C.E. | I.H.V. |
|---|---|---|---|---|
| SEdit | 135 | 16956 | 0.09 | 93% |
| Fractality[3] | 240 | 19042 | 0.86 | 25% |
| HSQLdb | 283 | 61633 | 0.37 | 72% |
| Jasper Reports | 316 | 99540 | 0 | 100% |
| JBPM | 366 | 133590 | 0 | 100% |
| Manta-Ray | 384 | 132298 | 0.12 | 89% |
| Blue Marine | 468 | 194991 | 0.13 | 89% |
| Jboss | 4244 | 17619836 | 0.02 | 98% |

3  Not on sourceforge, the parser is included to show that high configuration efficiency systems are indeed possible, which might otherwise have been questionable.

|        | Num. nodes | P.S.C.    | C.E. | I.H.V. |
|--------|-----------|-----------|------|--------|
| Eclipse | 39114    | 933916300 | 0.4  | 61%    |

*Table 2: Sample systems and associated metrics*

The systems depicted in table 2 present too small a sample for us to draw definitive conclusions (a broader, empirical survey is being researched). We note some casual observations, however:

- Both JasperReports and JPBM have made all their classes public, thus attaining minimum configuration efficiency.

- MantaRay has more classes than JBPM and yet expresses a lower P.S.C.; MantaRay achieved this through its higher configuration efficiency.

- Despite the astronomically high P.S.C., Eclipse has one of the highest configuration efficiencies of the selection.

- Only one system resides above the stable equilibrium belt, though it is stressed that these systems may employ sophisticated encapsulation strategies that elude the tool's simple parsing, so their configuration efficiencies could be higher.

Also, though not apparent from the statistics shown, Jboss re-encapsulated to its theoretical minimum P.S.C. given its current number of classes and regional information hiding violation, would reduce its current P.S.C. from 17 million to 1 million.

As a final note on the utility of configuration efficiency, consider that to minimise P.S.C. the number of program units in a subsystem should be proportional to the total number of program units (theorem 1.15). It is impractical, however, to redistribute program units during software development each time the number of program units change, so the common programming practice is to enforce a maximum number of program units per subsystem. This proves to be a good strategy for managing P.S.C.; there is a cost, but the configuration efficiency remains approximately scale invariant for a wide range of systems, a desirable quality in any encapsulation and information hiding strategy. Figure 24, for example, shows a graph of a system of increasing number of program units and its corresponding configuration efficiency (which levels off at 0.91), the constraint being that system is allowed to have at most ten program units per subsystem (again, $p=1$).

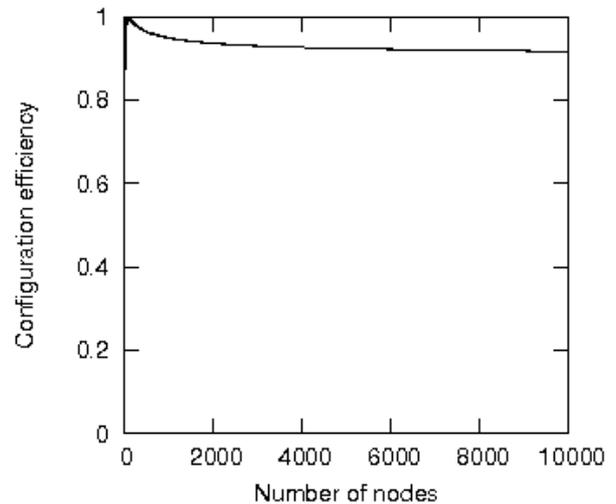

*Figure 24: Configuration efficiency with ten program units per subsystem*

### 6.6. Interpretations

So far we have discussed the encapsulation of program units within subsystems. We shall now generalise these concepts by taking advantage of the terminology of graph theory.

We shall introduce the term, "Encapsulated region," loosely defined as a subgraph containing nodes each of which is distinguishable by membership of that encapsulated region. Essentially, the subsystems discussed so far are encapsulated regions and the program units inside these subsystems are the nodes within the encapsulated regions. An encapsulated graph, then, is an any graph whose every node resides within an encapsulated region.

The mathematics of encapsulation theory are independent of computer programming and dependent only on the underlying graph theory; that such a theory is relevant to computer programming is due to our ability to view a collection of program units residing within subsystems as a graph of nodes residing within encapsulated regions.

It therefore does not seem unreasonable to take this view at least one stage deeper for just as a subsystem is a graph of nodes so too is a program unit a graph of functions.

Thus we have another encapsulated graph and all that has been said so far about program units within subsystems also applies to functions within program units, where functions may be considered nodes and program units may be considered encapsulated regions.

In this case, information hiding is the restricting of forming dependencies on any particular function within a program unit. To take a Java example again, where a program unit is a class and a function is method, the following class has two information-hiding violations (methods `go()` and `stop()`) because both functions can be called by classes outside the `Car` class:

```
class Car {
        public void go() {…};
        void stop() {…};
        private void alarm() {…};
}
```

Encapsulation theory can therefore answer questions such as: given 100 functions and a requirement for just two functions to be visible outside each program unit, what number of program units will minimise the P.S.C. of this system? The answer, by the second law again, is:

$$r_{min} = \sqrt{\frac{n}{p}} = \sqrt{\frac{100}{2}} = \sqrt{50} \approx 7$$

It may also be reasonable to take this view yet another stage deeper still as there is a long-standing tradition of viewing even lines of code as a graph. In his landmark paper on cyclomatic complexity [9], Thomas McCabe describes source code where, "Each node in the graph corresponds to a block of code in the program where the flow is sequential and the arcs correspond to branches taken in the program."

We may adopt McCabe's view (and thus risk similar critique [1]), establishing a third encapsulated graph whereby blocks of sequential program flow are considered nodes and functions are considered encapsulated regions. This view is particularly attractive in that most programming languages allow only a single point-of-entry into a function, and thus we may consider the first block of sequential flow to be the only information-hiding violation of any function ($p=1$).

We thus arrive at the encapsulated graph stack, see figure 25.

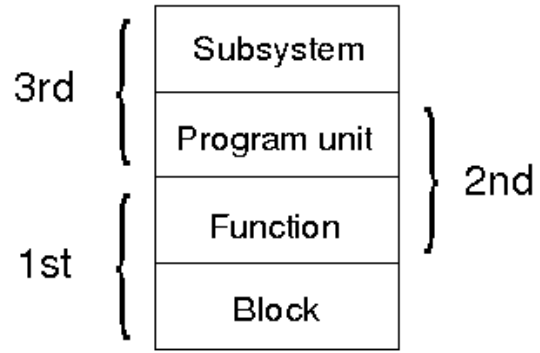

*Figure 25: Encapsulated graph stack*

In figure 25, we see the graph stack of the non-hierarchical encapsulation context. The first[4] is the block/function graph, where blocks are encapsulated into functions; the second is the function/program unit graph, where functions are encapsulated into program units; the third is the program unit/subsystem graph, where program units are encapsulated into subsystems.

Given that, usually, $p=1$ for the first graph, there is higher probability that software at the first graph will reside above the stable equilibrium belt than software at the higher graphs.

At the second graph, we note that there is a tendency to make functions public solely that they may be unit tested, in the belief that this has no consequence at the third graph given that the parent program units remain private within their subsystems. Encapsulation theory shows, however, that although increasing $p$ at the second graph decreases the number of subsystems needed at the third graph, the sum of the P.S.C.s of the two graphs increases with increased $p$ at the second graph.

### 6.7. New information hiding

So far we have considered only the non-hierarchical encapsulation context, which has a proven lower limit to its P.S.C. (theorem 1.14).

---

4    It is conceivable that subunits of lines of code may be encapsulated into lines of code themselves, thus giving rise to a zeroth encapsulated graph; this graph is not investigated in this paper.

Encapsulation contexts are defined by the type of information hiding they employ; we shall now investigate whether it is possible to attain even lower P.S.C. in a different encapsulation context. We shall first rename the information hiding discussed so far, "Absolute information hiding."

The new information hiding needed to further reduce P.S.C. is, "Relative information hiding."

According to both types of information hiding, a program unit defined as information hidden within its subsystem is forbidden to receive dependencies from outside that subsystem. This is inviolable. It is with respect to, "Public," program units that the two types differ.

According to absolute information hiding, a program unit defined as visible outside its subsystem is allowed to receive dependencies from all program units in all other subsystems. This, for example, is the standard treatment in Java, where a class declared as `public` is visible to all classes in all other packages.

According to relative information hiding, however, a program unit defined as visible outside its subsystem is allowed to receive dependencies only from those program units in other subsystems that enjoy a special relationship with the target subsystem.

What this, "Special relationship," means is usually defined by a system's software architecture, often imposed by consensual and extra-linguistic design rule. Java does not (at least, again, in advance of the introduction of super-packages) support relative information hiding but a Java system may, simply by specifying appropriate guidelines to programmers.

A common design rule used to establish relative information hiding, for instance, is the imposition of software layers, a layer usually being a collection of subsystems. This layer is then defined in relation to other software layers precisely so that relationships between layers can be controlled, a typical rule being of the form, "Lower layers may not depend on the program units of higher layers," where, "Lower," and, "Higher," refer either to technical abstraction, appropriate orientation. See figure 26.

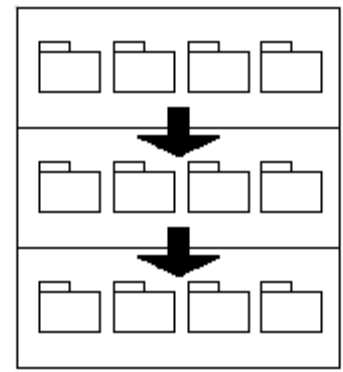

*Figure 26: Software layers*

In figure 26, we see three layers, each holding four subsystems; the arrows indicate that the dependencies between layers are allowed only from higher to lower layers. Though usually called, "Layering," we shall call this a, "One-dimensional hierarchical encapsulation context."

We shall briefly analyse this and the two-dimensional encapsulation hierarchical context to which relative information hiding gives rise, though note that, as opposed to absolute information hiding – which essentially takes only one form – there are endless variations of relative information hiding, and the mathematics of even the simplest variations are not well understood.

## 7. One-dimensional hierarchical encapsulation

In our brief overview of the one-dimensional hierarchical encapsulation context, we make two assumptions.

Firstly, we shall assume that layers do not contain other layers.

Secondly, we shall assume that layers do not admit differential information hiding of their encapsulated subsystems: all subsystems of a layer are visible to client layers without any being able to declare themselves as visible only to other subsystems within the layer itself.

## 7.1. The fixed-system experiment

Again, we shall perform the fixed-system experiment to investigate whether the P.S.C. of a specific system is minimised by its being uniformly distributed.

Consider twelve subsystems distributed over three layers. We shall begin with ten subsystems in layer 1 (the bottom layer) and one subsystem in each of the other two layers. Then we shall put nine subsystems in layer 1, two subsystems in layer 2 and one subsystem in layer 1, etc..

We shall perform this experiment using a computer simulation rather than drawing out the dependencies by hand. The results are shown in figure 27.

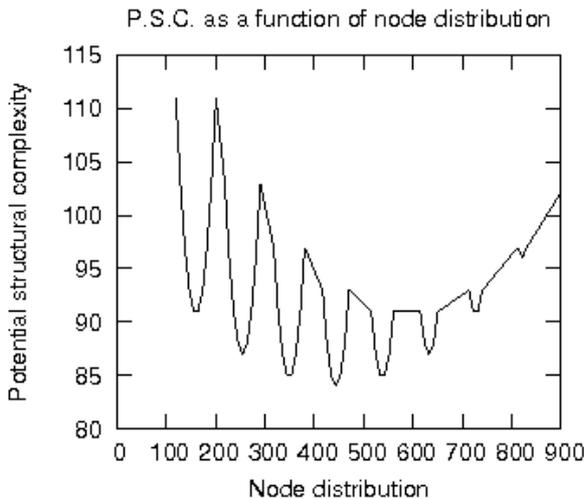

Figure 27: Distributing 12 subsystems over 3 layers

Figure 27 is slightly difficult to interpret but the data confirm that the system's P.S.C. is minimised when there are four subsystems in each layer (hence the minimum at 4-4-4), that is, when the subsystem are uniformly distributed overall the layers.

Although it has not been proven that uniform distribution minimises all one-dimensional hierarchical encapsulated systems, we shall take this experiment as evidence to support the conjecture.

## 7.2. The varied-region experiment

We shall perform our second experiment, the varied-region experiment, to attempt to establish whether there exists a system configuration (in this case, a particular number of layers) that minimises the P.S.C. of a system.

Consider again a system of 100 program units. We shall divide this system up into an increasing number of layers, such that the nodes will be uniformly distributed over any number of subsystems and these subsystems will be uniformly distributed over the layers.

First, we shall examine a system of one layer. Within this one layer, we shall examine all configurations of the 100 nodes distributed over 1 to 100 subsystems, and we shall record the configuration with the lowest P.S.C..

Second, we shall examine a system of two layers. Within these two layers, we shall examine all configurations of the 100 nodes distributed over 1 to 100 subsystems, and we shall record the configuration with the lowest P.S.C..

Next, we shall examine a system of three layers, etc., all the way to a system with 100 layers, with one subsystem per layer and one program unit per subsystem.

Each subsystem will have a regional information hiding violation of 1 ($p=1$). Program units of subsystems in each layer will be able to form dependences towards subsystems in the same layer, and all layers below. Program units in layer 3, for example, will be able to form dependencies towards program units in other subsystems in layer 3, and on those in layers 2 and 1.

The result of this minimum P.S.C. plotted against increasing numbers of layers is shown in figure 28.

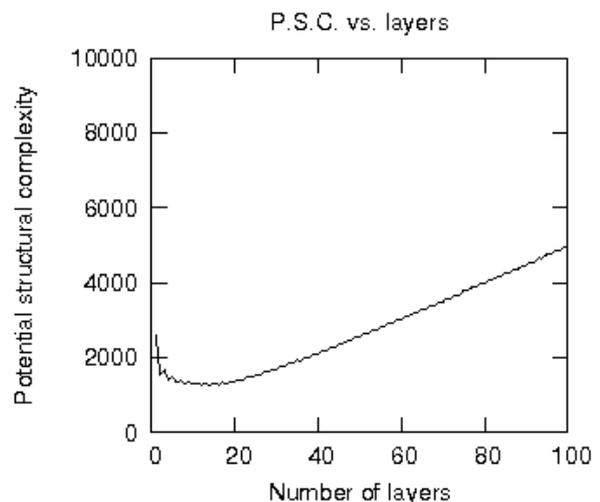

Figure 28: Distributing 100 program units over 1-

*100 layers*

Figure 28 suggests that there is indeed a number of layers which – when program units and subsystems are uniform distributed – minimises the P.S.C. of the system.

The number of layers is, of course, a parameter with which we were not confronted when we examined the non-hierarchical encapsulation context. Given here for completeness and without proof is the equation for the P.S.C. of such a one-dimensional hierarchical encapsulation:

$$s(G) = n\left(\frac{n}{r} - 1\right) + (r_L^2(d+1)\left(L - \frac{d}{2}\right) - r)\frac{pn}{r}$$

Where:

*n = number of program units*

*r = number of subsystems*

*L = number of layers*

*$r_L$ = number of subsystems per layer*

*d = dependency layer penetration*

*p = information hiding violation per subsystem*

### 7.3. The system-growth experiment

In this experiment, again, we shall examine all systems composed of one to one hundred program units. For each system, we shall find the configuration with the minimum P.S.C. such that all program units are uniformly distributed over subsystems and all subsystems are uniformly distributed over layers.

As in the previous experiment, each subsystem will have an information hiding violation of 1 (*p=1*) and program units of subsystems in each layer will be able to form dependences towards subsystems in the same layer, and all layers below.

We shall then plot this minimum P.S.C. as a function of increasing numbers of program units. This is shown in figure 29, plotted against the same results already shown for the unencapsulated and one-dimensional hierarchical encapsulated contexts so far encountered (in figure 16).

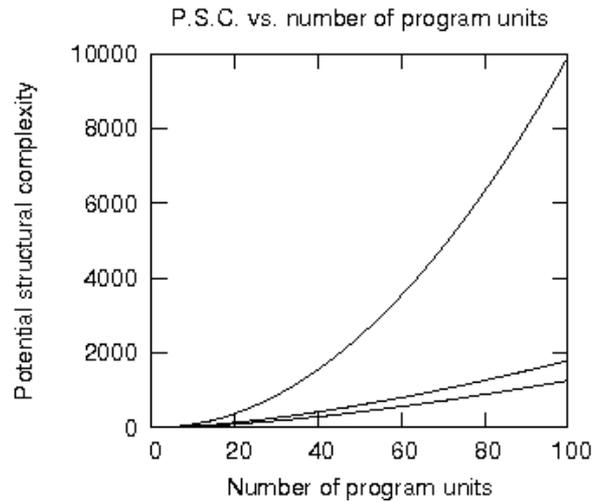

*Figure 29: Unencapsulated, non-hierarchical encapsulated and layered P.S.C. responses*

Figure 29 shows the minimum P.S.C. of systems containing 1 to 100 program units. The upper line is the unencapsulated system; the next line down is the non-hierarchical encapsulation context; the lowest line shows the one-dimensional hierarchical encapsulation context. As is evident, the layering of one-dimensional hierarchical encapsulation appears to reduce the minimum P.S.C. even beyond a non-hierarchical encapsulation context.

### 8. Two-dimensional hierarchical encapsulation

The one-dimensional hierarchical encapsulation context is one-dimensional in that only a single degree of restriction exists in forbidding dependencies between layers; essentially, a layer is either, "Above," or, "Below," another layer.

A simple two-dimensional hierarchical encapsulation context is achieved by taking the system described in figure 26 once again and adding the design rule that, for example, "Subsystems may not form dependencies on subsystems to their right." This effectively carves the figure into a two-dimensional plane, whereby the dependencies allowable are determined by a subsystems position in the x-axis (up or down the graphic) and the y-axis (position across the graphic).

Here, however, we hit a limitation of the layer as a second-class citizen of the software construct: simply put, whereas the imposition of dependency-restriction upon a collection of subsystems seems reasonable, programmers tend to baulk at the arbitrariness of assigning meaning to sub-collections of subsystems within layers.

We shall, therefore, investigate the more robust concept of a two-dimensional hierarchy based on recursively-defined subsystems. To do so, we must first present some simple graphical conventions. Consider the three subsystems shown in figure 30.

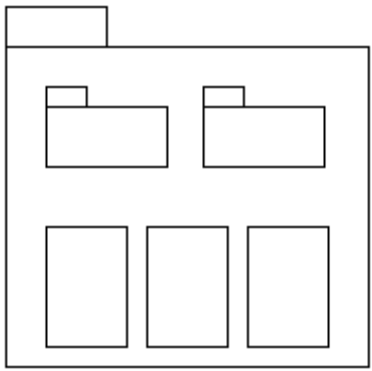

*Figure 30: A subsystem containing two other subsystems*

Figure 30 shows a large subsystem graphic enclosing two smaller subsystem graphics and three program unit graphics. This represents the larger subsystem's containing two subsystems and three program units. The larger subsystem is said to be the parent of the two smaller subsystems; the two smaller subsystems are said to be the children of the larger subsystem. If we concentrate solely on the subsystem graphics and ignore the program units, this containment is often drawn as a tree showing the two child subsystems above the parent subsystem, as in figure 31.

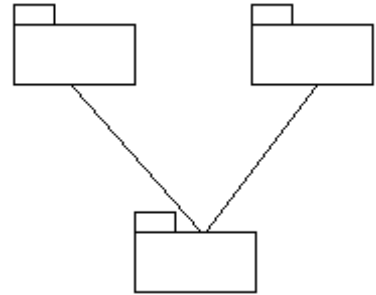

*Figure 31: A subsystem containing two other subsystems*

Figure 31 shows the parent subsystem on the bottom (in the root position, to extend the tree metaphor), and its two children on top. As well as the parent-child relationship, we can also define the peer-peer relationship: two subsystems are peers when the share the same immediate parent. Thus the two child subsystems in figure 31 are peer subsystems.

Finally, the span-of-control is defined as the number of children that a parental subsystem has.

Given these simple terms, we may then assess the efficacy with which a two-dimensional hierarchy manages P.S.C. once we define the governing dependency constraints. As previously, there is no limit to the number of constrains that we might employ; we could, for example, stipulate that, "No dependencies may be formed between peers," or that, "No dependencies may be formed towards parents;" both design rules would yield interesting results; but we shall not consider these here.

We shall consider the design rule, "Dependencies may not form towards children."[5] Thus, child subsystems are allowed to form dependencies towards peer subsystems, towards their parents, towards peers of their parents, towards parents" parents, towards peers of parents" parents, etc.. This rule forbids a program unit to form a dependency towards subsystems encapsulated within its own subsystem.

For example, consider the system depicted in figure 32.

---

5   As employed in *The Fractal Class Composition*, see author's website.

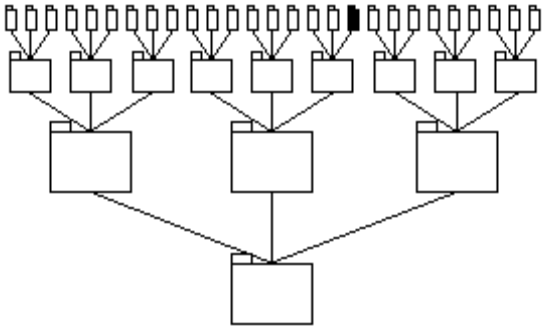

*Figure 32: A 2-d hierarchical encapsulation context with a span-of-control of 3*

Figure 32 shows a system of 40 subsystems arranged in a two-dimensional hierarchical encapsulation context. The span-of-control is three: that is, each subsystem contains three other subsystems.

Consider the shaded subsystem of figure 32, lying amid the uppermost (i.e., deepest recursively defined) subsystems: how many subsystems of the rest of the system can it see? The answer is shown in figure 33, whereby each visible subsystem is coloured black.

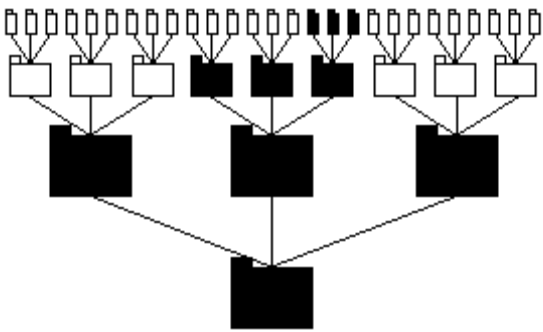

*Figure 33: A 2-d hierarchical system showing visibility from a leaf subsystem*

Figure 33 shows that the subsystem highlighted in figure 32 can see three subsystems on each level up from the root subsystem; in total, it can see only 10 of the system's 40 subsystems.

Computer simulation shows that, in fact, a span-of-control of two offers the most effective relative information hiding. Given here for completeness and without proof is the equation for the P.S.C. of a two-dimensional hierarchical encapsulation context:

$$s(G) = n\left(\frac{n}{r} - 1\right) + p\sum_{i=1}^{k} ib^{i+1}$$

Where:

*n = number of program units*

*r = number of subsystems*

*k = number of levels*

*b = span-of-control*

*p = information hiding violation per subsystem*

### 8.1. The fixed-system experiment

This experiment, to attempt to establish whether the P.S.C. of a two-dimensional hierarchical encapsulation context is minimised by the uniform distribution of program units and subsystems per subsystem, is not performed here and is the subject of on-going research.

### 8.2. The varied-region experiment

As before, let us perform our second experiment, the varied-region experiment, to try attempt to establish whether there exists a system configuration that minimises the P.S.C. of a two-dimensional hierarchical encapsulation context.

Let us take 100 program units and encapsulate them in increasing numbers of subsystems, each with an information hiding violation of one (*p=1*), starting with all program units in one subsystem.

We shall then distribute the 100 program units over two subsystems, whereby one subsystem will be the parent of the other and the parent will not be allowed to form any dependencies towards program units within the child subsystem.

Next, we shall distribute the program units over three subsystems, one being the parent of the other two. The fourth subsystem will then be the child of one of the children, and so on, all the while forbidding dependency formation towards child subsystems.

Thus we shall form a two-dimensional hierarchy of subsystems with a span-of-control of two and we shall record the P.S.C. of each configuration. The results are shown in figure 34.

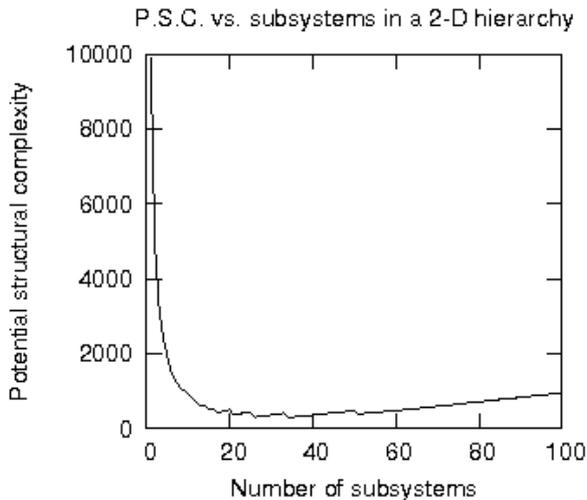

*Figure 34: Distributing 100 program units over subsystems in a 2-D hierarchy*

Figure 34 suggests that the P.S.C. does indeed fall and then rise again, though not as clearly as with previous experiments (in fact, as the number of subsystems grows in a two-dimensional hierarchy, the P.S.C. is quantised into ranges as new levels are created to accommodate the growing number of subsystems: hence the jittery nature of the graph).

Regardless of this somewhat unusual-looking result, the experiment at least suggests that, for any given system of $n$ program units, the system's P.S.C. is minimised for some uniform distribution when those subsystems form a two-dimensional hierarchical encapsulation context. (Again, this is not proven.)

Thus we can proceed with our third experiment, to attempt to examine how such systems of increasing numbers of program units express this P.S.C. minimum.

## 8.3. The system-growth experiment

In this experiment we shall examine all systems composed of one to one hundred program units. For each system, we shall find the configuration with the minimum P.S.C. such that all program units are uniformly distributed over subsystems and all subsystems are uniformly distributed over a two-dimensional hierarchical encapsulation context with a span-of-control of two.

As in the previous experiment, each subsystem will have an information hiding violation of one ($p=1$) and the dependency-formation towards child subsystems will be forbidden.

We shall then plot this minimum P.S.C. as a function of increasing numbers of program units. This is shown in figure 35, plotted against the same results already collected for this experiment in other contexts.

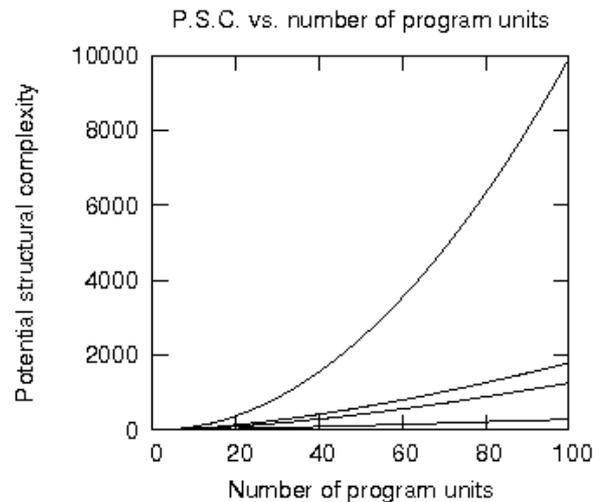

*Figure 35: Unencapsulated, non-hierarchical, 1-D and 2-D hierarchical encapsulation context P.S.C. responses*

Figure 35 shows the minimum P.S.C. of systems containing 1 to 100 program units. The upper line is the unencapsulated system; the next line down is the non-hierarchical encapsulation context; the third line shows the one-dimensional hierarchical encapsulation context; the lowest line shows the P.S.C. expressed by our two-dimensional hierarchical encapsulation context.

The two-dimensional hierarchical encapsulation context suggests a significant improvement over other contenders.

## 9. CONCLUSIONS

Information hiding and encapsulation have long been proposed as useful tools with which to reduce program complexity. This paper has proposed the concept of potential structural complexity – comprising both information hiding and encapsulation – as means to mathematically to explore this claim.

The three laws of encapsulation are formulated based on four quantities of a system:

- $s(G)$ – the P.S.C. function.
- $n$ – number of program units
- $r$ – number of subsystems
- $p$ – number of regional information hiding violations (i.e., number of public program units per subsystem).

The three laws are:

1. The maximum P.S.C. of a system is given by the equation (theorem 1.11):

$$s(G) = n(n - 1)$$

2. The minimum P.S.C. of a uniformly distributed system is expressed when the number of its subsystems is given by (theorem 1.12):

$$r = \sqrt{\frac{n}{p}}$$

3. The maximum P.S.C. of a uniformly distributed system is expressed when the number of its subsystems is given by (theorem 1.13):

$$r = \frac{n}{p}$$

The adage is appropriate once more, "Exploit encapsulation and information hiding optimally."

Finally, the mathematics presented here are not restricted to digital physics but apply to any system that can be readily interpreted as an encapsulated graph. A manager wishing to organise 20 people into teams with one team-leader in each team, such that each team may only communicate with the team-leader of the other teams, will find that communication effort will be minimised by the application of the second law of encapsulation (giving four teams of five members each).

## 10. Related work

Coupling and cohesion are introduced in [4]. External P.S.C. could be viewed as potential coupling; it is more problematic to view internal P.S.C. as a form of potential cohesion.

In that this paper proposes a graph theoretic optimal size for software system components ("the Goldilocks conjecture"), it proposes equation forms similar to those proposed in [8], which seeks to find an optimised size in terms of defect densities. That paper, however, was then critiqued in [7] , casting doubt on the assumptions made to yield those equations (the same critique cannot be levelled at P.S.C.).

## 11. APPENDIX A

### 11.1. DEFINITIONS

The theorems described proposed here are based on graph theory, so we shall build upon the accepted definitions of graph theory beginning with the graph itself. (Note that definitions and theorems presented here govern the non-hierarchical encapsulation context only.)

[D1] A directed graph $G$ is an ordered pair $G := (V,E)$ with:

- $V$ is a set whose elements are called nodes.
- $E$ is a set of ordered pairs of nodes called directed edges.
- The node from which an edge originates is called the tail; the node on which the edge terminates is called the head.

In this paper all graphs are assumed to be directed graphs.

[D2] The order of graph $G$ is the number of nodes in that graph, written $\lVert (G) \rVert$ .

[D3] A fully-connected graph is a graph whose every node has an edge towards every other node.

[D4] Two nodes of a directed graph are said to be consecutive if there exist two edges between them.

[D5] A labelled graph is a graph whose nodes are distinguished by labels or names.

[D6] Graph $G:=(V,E)$ contains a subgraph $K:=(W,B)$ iff:

(i)  $W \subseteq V \wedge B \subseteq E$

Note that a subgraph is also a graph.

[D7] An encapsulated region of a graph is a labelled subgraph whose nodes are exclusively labelled so that they form a set disjoint from all other encapsulated regions of the graph. Nodes of an encapsulated region are said to be internal to that encapsulated region; nodes not of an encapsulated region are said to be external to that region.

[D8] An encapsulated graph is a graph that contains one or more encapsulated regions. An encapsulated graph must contain no nodes existing outside any encapsulated region. An unencapsulated graph is a graph that contains no encapsulated regions. Thus given graph $G$ with $r$ encapsulated regions, where the $i^{th}$ encapsulated region is called $K_i$, the order of $G$ is the sum of the orders of $K_i$ or:

$$|G| = \sum_{i=1}^{r} |K_i|$$

Note that the order of an encapsulated region is the number of nodes immediately contained, not including the number of contained subsystems.

[D9] A node within an encapsulated region is information-hidden when nodes outside that encapsulated region are forbidden from forming a directed edge towards it. A node within an encapsulated region is an information-hiding violating node when nodes outside that encapsulated region are allowed to form a directed edge towards it. Note that an information-hidden node within an encapsulated region may still form a directed edge towards any information-hiding violating node outside that region.

[D10] All nodes within an encapsulated region are information-hiding violating nodes to one another, that is, they may all form directed edges towards one another.

[D11] The information-hiding violation function of encapsulated graph $G$, written $h(G)$, is the function that maps the information-hiding violating nodes of that graph to their own set. The information-hiding violation of graph $G$ is the cardinality of its information-hiding violation

function, $|h(G)|$. The information-hiding violation of encapsulated region $K$ is the cardinaltiy of its information-hiding violation function, $|h(K)|$. The information hiding violation of encapsulated region $K$ is also called the regional information hiding violation.

[D12] The information-hiding violation of a graph is the sum of the information-hiding violations of its encapsulated regions, or:

$$|h(G)| = \sum_{i=1}^{r} |h(K_i)|$$

[D13] A maximally-connected graph is an encapsulated graph whose every node has a directed edge towards every other accessible node. Note that maximal connectedness respects information-hiding, thus any two nodes may share two edges (i.e., the nodes are consecutive), one edge (i.e., one node is information-hidden within an encapsulated region) or zero edges (i.e., both nodes are information-hidden within two different encapsulation regions).

[D14] The structural complexity of a graph is the number of edges of that graph. Graphs are said to, "Express," structural complexity.

[D15] The internal structural complexity of an encapsulated region is the structural complexity expressed by edges whose heads and tails lie within that encapsulated region (the edges form within the encapsulated region).

[D16] The external structural complexity of an encapsulated region is the structural complexity expressed by edges whose tails lie within that encapsulated region but whose heads do not (the edges originate within encapsulated region but terminate outside it).

[D17] The structural complexity of an encapsulated region is the sum of that region's internal and external structural complexities. The structural complexity of a graph is the sum of the structural complexities of all its encapsulated regions.

[D18] The potential structural complexity of graph $G$, written $s(G)$, is the number of edges of that graph if it were maximally-connected. The potential structural complexity of graph $G$ is the sum of the potential structural complexities of all its encapsulated regions, or:

$$s(G) = \sum_{i=1}^{r} s(K_i)$$

[D19] The internal potential structural complexity of encapsulated region $K$, written $s_{in}(K)$, is the internal structural complexity of that region if it were part of maximally-connected graph. Given graph $G$ whose $i^{th}$ encapsulated region is $K_i$, the internal potential structural complexity of graph $G$, written $s_{in}(G)$, is the sum of the internal potential structural complexities of $K_i$, or:

$$s_{in}(G) = \sum_{i=1}^{r} s_{in}(K_i)$$

[D20] The external potential structural complexity of encapsulated region $K$, written $s_{ex}(K)$, is the external structural complexity of that encapsulated region if it were part of a maximally-connected graph. Given graph $G$ whose $i^{th}$ encapsulated region is $K_i$, the external potential structural complexity of graph $G$, written $s_{ex}(G)$, is the sum of the external potential structural complexities of $K_i$, or:

$$s_{ex}(G) = \sum_{i=1}^{r} s_{ex}(K_i)$$

[D20] The potential structural complexity of encapsulated region $K$, written $s(K)$, is the sum of the internal and external potential structural complexities of $K$, or:

$$s(K) = s_{in}(K) + s_{ex}(K)$$

The potential structural complexity of encapsulated graph $G$, written $s(G)$, is the sum of the internal and external potential structural complexities of $G$, or:

$$s(G) = s_{in}(G) + s_{ex}(G)$$

[D21] An encapsulated graph is uniformly distributed when each of its encapsulated regions contains the same number of nodes and has an equal information-hiding violation. Thus for graph $G$ of $n$ nodes and $r$, $G$'s being uniformly distributed implies:

(i)  $|K_1| = |K_2| = ... = |K_r| = \dfrac{n}{r}$

(ii)  $|h(K_1)| = |h(K_2)| = ... = |h(K_r)| = p$

(iii)  $p = |h(K_i)| = \dfrac{|h(G)|}{r}$

[D22] Two encapsulated graphs are equivalent if they have the same number of nodes, the same number of encapsulated regions and the same regional information hiding violation.

## 11.2. THEOREMS

### Theorem 1.1: the first law of encapsulation.

Given an unencapsulated graph $G$ of $n$ nodes, the potential structural complexity $s(G)$ is given by:

$$s(G) = n(n-1)$$

*Proof:*

To count the maximum possible number of edges between the $n$ nodes is to perform the selection of an order of arrangements of 2 nodes, without repetition, selected from the $n$ distinct nodes. That is, by definition, the permutation of n nodes taken 2 at a time:

$$s(G) = \frac{n!}{(n-2)!} = \frac{n \cdot (n-1) \cdot (n-2)!}{(n-2)!} = n(n-1)$$

*QED*

### Theorem 1.2.

Given an encapsulated region $K_i$ of $|K_i|$ nodes, the internal potential structural complexity $s_{in}(K_i)$ is given by:

$$s_{in}(K_i) = |K_i|(|K_i| - 1)$$

Proof:

The internal potential structural complexity of $K_i$ by definition has no contribution to the structural complexity of the rest of the graph; its internal potential structural complexity is expressed entirely by its encapsulated nodes. As each node within the encapsulated region may be consecutive with all others nodes in the region, then number of edges that may be formed is given by theorem 1.1:

$$s(G) = n(n-1)$$

Substituting $K_i$ for $G$, and $|K_i|$ for $n$ gives:

$$s_{in}(K_i) = \left|K_i\right|\left(\left|K_i\right|-1\right)$$

<div align="right"><em>QED</em></div>

## Theorem 1.3.

Given an encapsulated graph $G$ of $r$ encapsulated regions, of information hiding violation $h(G)$ and given that the $i^{th}$ encapsulated region $K_i$ contains $\left|K_i\right|$ nodes and has an information hiding violation of $h(K)$, the number of external nodes, $x$, towards which $K_i$ may direct edges towards is given by:

$$x = \left(\left|h(G)\right|-\left|h(K_i)\right|\right)$$

*Proof:*

By definition, the number of nodes external to $K_i$ is the sum of the numbers of nodes in all encapsulated regions except $K_i$, or:

$$\text{num. nodes external to } K_i = \sum_{j=1 \wedge j \neq i}^{r} \left|K_j\right|$$

But nodes in $K_i$ may not form edges towards all these other nodes, as some may be information hidden; $K_i$ may only form edges towards information-hiding violating nodes.

By definition, the $x$ we seek is the number of information-hiding violating nodes external to $K_i$, which is the sum of the numbers of information-hiding violating nodes in all encapsulated regions except $K_i$, or:

$$x = \sum_{j=1 \wedge j \neq i}^{r} \left|h(K_j)\right|$$

By definition, the information hiding violation of the entire graph, $\left|h(G)\right|$, is the sum of the information hiding violations of all its encapsulated regions:

$$\left|h(G)\right| = \sum_{j=1}^{r} \left|h(K_j)\right|$$
$$= \left|h(K_1)\right|+\left|h(K_2)\right|+...+\left|h(K_r)\right|$$

If we consider the $i^{th}$ encapsulated region $K_i$ in this sequence, we can write:

$$\left|h(G)\right| =$$
$$\left|h(K_1)\right|+\left|h(K_2)\right|+...$$
$$+\left|h(K_{i-1})\right|+\left|h(K_i)\right|+\left|h(K_{i+1})\right|+...$$
$$+\left|h(K_r)\right|$$

And thus,

$$\left|h(K_1)\right|+\left|h(K_2)\right|+...$$
$$+\left|h(K_{i-1})\right|+\left|h(K_{i+1})\right|+...+\left|h(K_r)\right|$$
$$= \left|h(G)\right|-h(K_i)$$

But,

$$\left|h(K_1)\right|+\left|h(K_2)\right|+...$$
$$+\left|h(K_{i-1})\right|+\left|h(K_{i+1})\right|+...+\left|h(K_r)\right|$$
$$= \sum_{j=1 \wedge j \neq i}^{r} \left|h(K_j)\right| = x$$

Therefore,

$$x = \left(\left|h(G)\right|-\left|h(K_i)\right|\right)$$

<div align="right"><em>QED</em></div>

## Theorem 1.4.

Given an encapsulated graph $G$ of $r$ encapsulated regions, of information hiding violation $h(G)$ and given that the $i^{th}$ encapsulated region $K_i$ contains $\left|K_i\right|$ nodes and has an information hiding violation of $h(K_i)$, the external potential structural complexity $s_{ex}(K_i)$ of $K_i$ is given by:

$$s_{ex}(K_i) = \left|K_i\right|\left(\left|h(G)\right|-\left|h(K_i)\right|\right)$$

*Proof:*

By definition, the external potential structural complexity of an encapsulated region is the number of edges formed out of that encapsulated region were it maximally connected.

The number of edges formable out of encapsulated region $K_i$ is the number of tails (nodes within $K_i$) multiplied by the number of external heads (nodes towards which $K_i$ may direct edges).

But $K_i$ contains $\left|K_i\right|$ nodes and by theorem 1.3 the number of external nodes towards which $K_i$ may direct edges is given by:

$$\left(\left|h(G)\right|-\left|h(K_i)\right|\right)$$

Therefore the external potential structural complexity $s_{ex}(K_i)$ of $K_i$ is given by:

$$s_{ex}(K_i) = \left|K_i\right|\left(\left|h(G)\right|-\left|h(K_i)\right|\right)$$

*QED*

### Theorem 1.5.

Given a uniformly distributed, encapsulated graph $G$, of $n$ nodes and of $r$ encapsulated regions and given that the $i^{th}$ encapsulated region is $K_i$, the internal potential structural complexity of $K_i$, $s_{in}(K_i)$, is given by:

$$s_{in}(K_i) = \frac{n}{r}\left(\frac{n}{r}-1\right)$$

*Proof:*

By theorem 1.2:

$$s_{in}(K_i) = |K_i|\left(|K_i|-1\right) \quad (i)$$

By definition, the encapsulated regions of uniformly distributed system all contain the average number of nodes per encapsulated region, or:

$$|K_i| = \frac{n}{r} \quad (ii)$$

Substituting (ii) in (i) gives:

$$s_{in}(K_i) = \frac{n}{r}\left(\frac{n}{r}-1\right)$$

*QED*

### Theorem 1.6.

Given a uniformly distributed, encapsulated graph $G$, of $n$ nodes and of $r$ encapsulated regions and given that the $i^{th}$ encapsulated region is $K_i$, the internal potential structural complexity of $G$, $s_{in}(G)$, is given by:

$$s_{in}(G) = n\left(\frac{n}{r}-1\right)$$

*Proof:*

By definition, the internal potential structural complexity of $G$ is the sum of the internal structural complexities of all its encapsulated regions, or:

$$s_{in}(G) = \sum_{i=1}^{r} s_{in}(K_i) \quad (i)$$

By theorem 1.5:

$$s_{in}(K_i) = \frac{n}{r}\left(\frac{n}{r}-1\right) \quad (ii)$$

Substituting (ii) into (i) gives:

$$s_{in}(G) = \sum_{i=1}^{r} \frac{n}{r}\left(\frac{n}{r}-1\right)$$

$$= r\frac{n}{r}\left(\frac{n}{r}-1\right)$$

$$= n\left(\frac{n}{r}-1\right)$$

*QED*

### Theorem 1.7.

Given a uniformly distributed, encapsulated graph $G$, of $n$ nodes and of $r$ encapsulated regions, with an information hiding violation of $h(G)$ and given that the $i^{th}$ encapsulated region $K_i$ has an information hiding violation of $p$, the external potential structural complexity $s_{ex}(G)$ is given by:

$$s_{ex}(G) = n(r-1)p$$

*Proof:*

By definition, the external potential structural complexity of graph $G$ is the sum of the external potential structural complexities of all its encapsulated regions, or:

$$s_{ex}(G) = \sum_{i=1}^{r} s_{ex}(K_i) \quad (i)$$

By theorem 1.4:

$$s_{ex}(K_i) = |K_i|\left(|h(G)|-|h(K_i)|\right) \quad (ii)$$

Substituting (ii) into (i) gives:

$$s_{ex}(G) = \sum_{i=1}^{r} |K_i|\left(|h(G)|-|h(K_i)|\right) \quad (iii)$$

By the definition of a uniformly distributed graph, the number of nodes per encapsulated region is the same, that is,

$$|K_1| = |K_2| = \ldots = |K_r| = \frac{n}{r} \quad (iv)$$

Substituting (iv) into (iii) gives:

$$s_{ex}(G) = \sum_{i=1}^{r} \frac{n}{r}\left(|h(G)|-|h(K_i)|\right)$$

$$= \frac{n}{r}\sum_{i=1}^{r}\left(|h(G)|-|h(K_i)|\right)$$

$$= \frac{n}{r}\sum_{i=1}^{r}|h(G)| - \frac{n}{r}\sum_{i=1}^{r}|h(K_i)|$$

$$= \frac{n}{r}r|h(G)| - \frac{n}{r}\sum_{i=1}^{r}|h(K_i)|$$

$$= n|h(G)| - \frac{n}{r}\sum_{i=1}^{r}|h(K_i)| \quad \text{(v)}$$

By definition, the sum of the information hiding violations of all encapsulated regions is the information hiding violation of the entire graph, thus,

$$|h(G)| = \sum_{i=1}^{r}|h(K_i)| \quad \text{(vi)}$$

Substituting (vi) into (v) gives,

$$s_{ex}(G) = n|h(G)| - \frac{n}{r}|h(G)| \quad \text{(vii)}$$

By definition, $p$ is the information hiding violation per region,

$$p = \frac{|h(G)|}{r} \rightarrow |h(G)| = rp \quad \text{(viii)}$$

Substituting (viii) into (vii) gives:

$$s_{ex}(G) = nrp - \frac{n}{r}rp$$

$$= nrp - np$$

$$= n(rp - p)$$

$$= n(r-1)p$$

*QED*

### Theorem 1.8: The potential structural complexity theorem.

Given a uniformly distributed, encapsulated graph $G$, of $n$ nodes and of $r$ encapsulated regions, with an information hiding violation of $h(G)$ and given that the $i^{th}$ encapsulated region $K_i$ has an information hiding violation of $p$, the potential structural complexity $s(G)$ is given by:

$$s(G) = n\left(\frac{n}{r} - 1 + (r-1)p\right)$$

*Proof:*

By definition, the potential structural complexity of encapsulated graph $G$, written $s(G)$, is the sum of the internal and external potential

structural complexities of $G$, or:

$$s(G) = s_{in}(G) + s_{ex}(G) \quad \text{(i)}$$

By theorem 1.6 the internal potential structural complexity $s_{in}(G)$ is given by:

$$s_{in}(G) = n\left(\frac{n}{r} - 1\right) \quad \text{(ii)}$$

By theorem 1.7 the external potential structural complexity $s_{ex}(G)$ is given by:

$$s_{ex}(G) = n(r-1)p \quad \text{(iii)}$$

Substituting (iii) and (ii) into (i) gives:

$$s(G) = n\left(\frac{n}{r} - 1\right) + n(r-1)p$$

$$= n\left(\frac{n}{r} - 1 + (r-1)p\right)$$

*QED*

### Theorem 1.9.

Given an encapsulated graph $U$, not uniformly distributed, of $n$ nodes and given that each $i^{th}$ encapsulated region $K_i$ contains $|K_i|$ nodes, the sum of the deviations from mean numbers of nodes per encapsulated region is 0.

*Proof:*

Consider that the nodes within graph $U$ are not equally distributed, but that each encapsulated region contains $\left(\frac{n}{r} + d_i\right)$ nodes, where $\frac{n}{r}$ is the mean number of nodes per region and $d_i$ is some deviation from that mean, then we can see that:

$$|K_i| = \frac{n}{r} + d_i \quad \text{(i)}$$

As the number of nodes, $n$, is by definition the sum of the nodes in all encapsulated regions, then,

$$n = \sum_{i=1}^{r}|K_i| \quad \text{(ii)}$$

Substituting (ii) into (i) gives,

$$n = \sum_{i=1}^{r}\left(\frac{n}{r} + d_i\right) = \sum_{i=1}^{r}\frac{n}{r} + \sum_{i=1}^{r}d_i$$

$$= r\frac{n}{r} + \sum_{i=1}^{r} d_i$$

$$n = n + \sum_{i=1}^{r} d_i$$

And,

$$\sum_{i=1}^{r} d_i = 0$$

*QED*

### Theorem 1.10.

Given an encapsulated graph *U*, not uniformly distributed, of information hiding violation *h(U)*, and given that each $i^{th}$ encapsulated region $K_i$ contains $|K_i|$ nodes and has the same information hiding violation of *p*, the external potential structural complexity is independent of distribution of nodes within those encapsulated regions.

*Proof:*

We shall attempt to prove this theorem by showing that equation for the external potential structural complexity of a graph whose nodes are not uniformly distributed is the same as the equation for the external potential structural complexity of a graph whose nodes are uniformly distributed; if the two equations are the same, then the external potential structural complexity of a graph is independent of node distribution.

By theorem 1.7 the external potential structural complexity, $s_{ex}(G)$, of a uniformly distributed graph is given by:

$$s_{ex}(G) = n(r-1)p \quad \text{(i)}$$

By definition, given encapsulated graph *U*, the external potential structural complexity of *s(U)* is given by:

$$s_{ex}(U) = \sum_{i=1}^{r} s_{ex}(K_i) \quad \text{(ii)}$$

By theorem 1.3,

$$s_{ex}(K_i) = |K_i|(|h(U)| - |h(K_i)|) \quad \text{(iii)}$$

Substituting (iii) into (ii) gives,

$$s_{ex}(U) = \sum_{i=1}^{r} |K_i|(|h(U)| - |h(K_i)|)$$

Also, given that each encapsulated region has the same information hiding violation, *p*, or:

$$|h(K_i)| = p$$

Substituting gives,

$$s_{ex}(U) = \sum_{i=1}^{r} |K_i|(|h(U)| - p) \quad \text{(i)}$$

If we now consider that the nodes within graph *U* are not equally distributed, but that each region contains $(\frac{n}{r} + d_i)$ nodes, where $\frac{n}{r}$ is the mean number of nodes per region and $d_i$ is some deviation from that mean. Thus,

$$|K_i| = \frac{n}{r} + d_i \quad \text{(ii)}$$

Substituting (ii) into (i) gives,

$$s_{ex}(U) = \sum_{i=1}^{r} (\frac{n}{r} + d_i)(|h(U)| - p)$$

As both $|h(U)|$ and *p* are independent of *i*, then we can write,

$$s_{ex}(U) = (|h(U)| - p) \sum_{i=1}^{r} (\frac{n}{r} + d_i)$$

$$= (|h(U)| - p) \sum_{i=1}^{r} \frac{n}{r} + (|h(U)| - p) \sum_{i=1}^{r} d_i \quad \text{(iii)}$$

Theorem 1.9 shows that:

$$\sum_{i=1}^{r} d_i = 0 \quad \text{(iv)}$$

Substituting (iv) into (iii) gives:

$$s_{ex}(U) = (|h(U)| - p) \sum_{i=1}^{r} \frac{n}{r}$$

$$= (|h(U)| - p) r\frac{n}{r}$$

$$= (|h(U)| - p)n \quad \text{(v)}$$

Even though the nodes of U are non-uniformly distributed over its encapsulated regions, each encapsulated region does have the same information hiding violation, so by definition:

$$|h(U)| = rp \quad \text{(vi)}$$

Substituting (vi) into (v) gives:

$$s_{ex}(U) = (rp - p)n$$

$$= n(r-1)p$$

This equation is the same as (i) which governs a system uniformly distributed and therefore the external potential structural complexity is independent of distribution of nodes within those encapsulated regions.

*QED*

### Theorem 1.11.

Given a uniformly distributed, encapsulated graph $G$, of $n$ nodes and of $r$ encapsulated regions, with each encapsulated region having an information hiding violation of $p$, there does not exist any graph, $U$, not uniformly distributed, of equal $n$, $r$ and $p$, with a lower potential structural complexity.

*Proof:*

We shall assume that this theorem is false and prove that this leads to a contradiction. Thus, consider a graph $U$, not uniformly distributed, of $n$ nodes and $r$ encapsulated regions, with each encapsulated region having an information-hiding violation of $p$; as $n$, $r$ and $p$ are the same for both $G$ and $U$, then the graphs differ only in how nodes are distributed across encapsulated regions: the nodes of $G$ are uniformly distributed, and the nodes of $U$ are not uniformly distributed.

The $i^{th}$ encapsulated region of $U$ is $J_i$ and this region contains $|J_i|$ nodes. By definition, the potential structural complexity of $J_i$ is given by:

$$s(J_i) = s_{in}(J_i) + s_{ex}(J_i)$$

By definition, the potential structural complexity of the entire graph $U$ is given by:

$$s(U) = \sum_{i=1}^{r} [s_{in}(J_i) + s_{ex}(J_i)]$$

If we take the $i^{th}$ encapsulated region of $G$ to be $K_i$, then by definition,

$$s(G) = \sum_{i=1}^{r} [s_{in}(K_i) + s_{ex}(K_i)]$$

Assuming that the theorem's premise is false implies assuming that the potential structural complexity of $G$ is greater than that of $U$, or:

$$s(G) > s(U)$$

And thus,

$$\sum_{i=1}^{r} [s_{in}(K_i) + s_{ex}(K_i)] > \sum_{i=1}^{r} [s_{in}(J_i) + s_{ex}(J_i)]$$

But as $G$ and $U$ differ only in the distribution of their nodes, by theorem 1.10 their external potential structural complexities must be equal. Hence,

$$\sum_{i=1}^{r} s_{in}(K_i) > \sum_{i=1}^{r} s_{in}(J_i) \quad \text{(i)}$$

By theorem 1.2, the internal potential structural complexity of $J_i$ can be written as:

$$s_{in}(J_i) = |J_i|(|J_i| - 1) \quad \text{(ii)}$$

Given that $U$ is not uniformly distributed and thus that the nodes within $U$'s encapsulated regions are not uniformly distributed, we can define that each region contains $(c + d_i)$ nodes, where $c$ is defined as the mean number of nodes per region and $d_i$ is some deviation from that mean, then:

$$|J_i| = c + d_i \quad \text{(iii)}$$

We also note that $U$'s not being uniformly distributed implies that $d_i$ is non-zero for some $i$, or:

$$\exists i : d_i \neq 0$$

This implies:

$$\sum_{i=1}^{r} d_i^2 > 0 \quad \text{(iv)}$$

Substituting (iii) into (ii) gives,

$$s_{in}(J_i) = (c + d_i)(c + d_i - 1) \quad \text{(v)}$$

The internal potential structural complexity of $K_i$ is also given by theorem 1.2 as,

$$s_{in}(K_i) = |K_i|(|K_i| - 1)$$

As $G$ is uniformly distributed, however, then each encapsulated region contains the same number of nodes, $c$, or:

$$|K_i| = c$$

Substituting this into (vi) gives,

$$s_{in}(K_i) = c(c - 1) \quad \text{(vii)}$$

Substituting into (vii) and (v) into (i) gives,

$$\sum_{i=1}^{r} c(c-1) > \sum_{i=1}^{r} (c+d_i)(c+d_i-1)$$

$$r\,c(c-1) > \sum_{i=1}^{r} (c^2 + cd_i - c + c\,d_i + d_i^2 - d_i)$$

$$r\,c(c-1) > \sum_{i=1}^{r} (c^2 + 2cd_i - c + d_i^2 - d_i)$$

$$r\,c(c-1) > rc^2 + 2c\sum_{i=1}^{r} d_i - rc + \sum_{i=1}^{r} d_i^2 - \sum_{i=1}^{r} d_i$$

Theorem 1.9 proves that:

$$\sum_{i=1}^{r} d_i = 0 \quad \text{(vi)}$$

Substituting (vi) gives,

$$r\,c(c-1) > rc^2 - rc + \sum_{i=1}^{r} d_i^2$$

$$r\,c(c-1) > r\,c(c-1) + \sum_{i=1}^{r} d_i^2$$

$$0 > \sum_{i=1}^{r} d_i^2$$

This conflicts with (iv) and thus the initial assumption – that there exists a graph, *U*, not uniformly distributed, whose potential structural complexity is lower than that of *G* – must be false.

*QED*

## Theorem 1.12: the second law of encapsulation.

Given a uniformly distributed, encapsulated graph *G*, of *n* nodes and of *r* encapsulated regions, with each encapsulated region having an information hiding violation of *p*, the number of encapsulated regions $r_{min}$ that minimises the graph's potential structural complexity is given by:

$$r_{min} = \sqrt{\frac{n}{p}}$$

*Proof:*

By theorem 1.8, the potential structural complexity *s(G)* is given by:

$$s(G) = n\left(\frac{n}{r} - 1 + (r-1)\,p\right)$$

To find the minimum potential structural complexity, differentiate with respect to r and set to zero.

$$\frac{\partial}{\partial r} s(G) = \frac{\partial}{\partial r}\left[n\left(\frac{n}{r} - 1 + (r-1)\,p\right)\right] = 0$$

$$= n\left[\frac{\partial}{\partial r}\left(\frac{n}{r} - 1 + (r-1)\,p\right)\right]$$

$$\frac{\partial}{\partial r}\left(\frac{n}{r} - 1 + rp - p\right) = 0$$

$$= \frac{-n}{r^2} + p$$

$$= -n + r^2\,p$$

$$r^2 p = n$$

$$r^2 = \frac{n}{p}$$

$$r_{min} = \sqrt{\frac{n}{p}}$$

*QED*

## Theorem 1.13: the third law of encapsulation.

Given a uniformly distributed, encapsulated graph *G*, of *n* nodes and of *r* encapsulated regions, with each encapsulated region having an information hiding violation of *p*, the number of encapsulated regions $r_h$ at which point the potential structural complexity reaches that of an equivalently-ordered unencapsulated graph is given by:

$$r_h = \frac{n}{p}$$

*Proof:*

To find where the potential structural complexity of *G* intersects that of the equivalently ordered unencapsulated system, equate the equations of both potential structural complexities and solve for *r*.

From theorem 1.1, the equation for the potential structural complexity of an

unencapsulated graph is:

$$s(G) = n(n-1)$$

From theorem 1.8, the equation for the potential structural complexity of an encapsulated graph uniformly distributed is:

$$s(G) = n\left(\frac{n}{r} - 1 + (r-1)p\right)$$

Equating the two gives,

$$n(n-1) = n\left(\frac{n}{r} - 1 + (r-1)p\right)$$

$$n - 1 = \frac{n}{r} - 1 + (r-1)p$$

$$n = \frac{n}{r} + rp - p$$

$$nr = n + pr^2 - pr$$

$$pr^2 - pr - nr + n = 0$$

$$pr^2 + (-p-n)r + n = 0$$

$$r = \frac{p + n \pm \sqrt{(-p-n)^2 - 4np}}{2p}$$

$$= \frac{p + n \pm \sqrt{p^2 + 2np + n^2 - 4np}}{2p}$$

$$= \frac{p + n \pm \sqrt{p^2 - 2np + n^2}}{2p}$$

$$= \frac{p + n \pm \sqrt{(p-n)^2}}{2p}$$

$$r = \frac{p + n \pm (p-n)}{2p}$$

If we take the first root,

$$r = \frac{p + n + (p-n)}{2p}$$

$$= \frac{p + n + p - n}{2p}$$

$$= \frac{2p}{2p}$$

$$r = 1$$

Take second root,

$$r = \frac{p + n - (p-n)}{2p}$$

$$= \frac{2n}{2p}$$

$$r_h = \frac{n}{p}$$

*QED*

### Theorem 1.14.

Given a uniformly distributed, encapsulated graph $G$, of $n$ nodes and of $r$ encapsulated regions, with each encapsulated region having an information hiding violation of $p$, the graph's minimum potential structural complexity is given by:

$$s_{min}(G) = n(2\sqrt{np} - 1 - p)$$

*Proof:*

From theorem 1.8, the potential structural complexity, $s(G)$, is given by:

$$s(G) = n\left(\frac{n}{r} - 1 + (r-1)p\right) \quad (i)$$

From theorem 1.12, the number of encapsulated regions, $r_{min}$, that minimises the graph's potential structural complexity is given by:

$$r_{min} = \sqrt{\frac{n}{p}} \quad (ii)$$

Substituting (ii) into (i) gives,

$$s(G) = n\left(\frac{n}{\sqrt{\frac{n}{p}}} - 1 + \left(\sqrt{\frac{n}{p}} - 1\right)p\right)$$

$$= n\left(n\sqrt{\frac{p}{n}} - 1 + p\sqrt{\frac{n}{p}} - p\right)$$

$$= n(\sqrt{np} - 1 + \sqrt{np} - p)$$

$$= n(2\sqrt{np} - 1 - p)$$

*QED*

### Theorem 1.15.

Given a uniformly distributed, encapsulated graph $G$, of $n$ nodes and of $r$

encapsulated regions, and given that the $i^{th}$ encapsulated region $K_i$ has an information hiding violation of $p$, the number of nodes in $K_i$ when the graph's potential structural complexity is minimised is given by:

$$|K_i| = \sqrt{np}$$

*Proof:*

As $G$ is evenly distributed then the number of nodes in $K_i$ is the average number of nodes per encapsulated region, or:

$$|K_i| = \frac{n}{r} \quad \text{(i)}$$

From theorem 1.12, the number of encapsulated regions $r_{min}$ that minimises the graph's potential structural complexity is given by:

$$r_{min} = \sqrt{\frac{n}{p}} \quad \text{(ii)}$$

Substituting (ii) into (i) gives:

$$|K_i| = \frac{n}{\sqrt{\dfrac{n}{p}}}$$

$$= n\sqrt{\frac{p}{n}}$$

$$= \sqrt{n^2 \frac{p}{n}}$$

$$= \sqrt{np}$$

*QED*

### Theorem 1.16.

Given a uniformly distributed, encapsulated graph $G$, of $n$ nodes, and given that the $i^{th}$ encapsulated region $K_i$ has an information hiding violation of $p$, the information hiding violation of $K_i$ when the graph's potential structural complexity is minimised is given by:

$$p = \frac{|K_i|^2}{n}$$

Proof:

From theorem 1.15, the number of nodes in $K_i$ when the graph's potential structural complexity is minimised is given by:

$$|K_i| = \sqrt{np}$$

Thus,

$$|K_i|^2 = np$$

$$p = \frac{|K_i|^2}{n}$$

*QED*